\documentclass[a4paper,11pt]{article}
\pdfoutput=1
\usepackage{jcappub}
\usepackage{bbold}
\usepackage[x11names]{xcolor}
\usepackage{colortbl,hhline}
\usepackage[english]{babel}
\usepackage[utf8]{inputenc}
\usepackage{amsmath}
\usepackage{color}
\usepackage{amsfonts}
\usepackage{graphicx}
\usepackage{amssymb}
\usepackage{eufrak}
\usepackage{etoolbox}
\usepackage{amsmath}
\usepackage{empheq}
\usepackage{cancel}
\usepackage[most]{tcolorbox}
\usepackage{float}              
\usepackage{subcaption}
\usepackage[normalem]{ulem}
\usepackage{xspace}
\usepackage{adjustbox}
\newtcbox{\mymath}[1][]{%
    nobeforeafter, math upper, tcbox raise base,
    enhanced, colframe=yellow!30!black,
    colback=yellow!30, boxrule=1pt,
    #1}

\def\be{\begin{equation}}
\def\ee{\end{equation}}
\def\bea{\begin{eqnarray}}
\def\eea{\end{eqnarray}}
\def\bean{\begin{eqnarray*}}
\def\eean{\end{eqnarray*}}

\newcommand{\HH}{\mathcal H}

\definecolor{bittersweet}{rgb}{1.0, 0.44, 0.37}

\renewcommand{\arraystretch}{1.8}

\newcommand{\MP}{M_{\rm Pl}}
\newcommand{\zetaf}{\zeta}

\newcommand{\lat}{{\mathrm i,\mathrm j,\mathrm k}}
\newcommand {\mf}{\mathrm}
\newcommand {\sn}{\mathrm n}
\newcommand {\gev}{\textit{gevolution}\xspace}
\newcommand {\kev}{{$k$-evolution}\xspace}
\newcommand{\class}{\texttt{CLASS}\xspace}
\newcommand{\hiclass}{\texttt{hi\_class}\xspace}

\usepackage{scalerel,tikz}
\usetikzlibrary{svg.path}
\definecolor{orcidlogocol}{HTML}{A6CE39}
\tikzset{orcidlogo/.pic={
 \fill[orcidlogocol] svg{M256,128c0,70.7-57.3,128-128,128C57.3,256,0,198.7,0,128C0,57.3,57.3,0,128,0C198.7,0,256,57.3,256,128z};
 \fill[white] svg{M86.3,186.2H70.9V79.1h15.4v48.4V186.2z}
 svg{M108.9,79.1h41.6c39.6,0,57,28.3,57,53.6c0,27.5-21.5,53.6-56.8,53.6h-41.8V79.1z M124.3,172.4h24.5c34.9,0,42.9-26.5,42.9-39.7c0-21.5-13.7-39.7-43.7-39.7h-23.7V172.4z}
 svg{M88.7,56.8c0,5.5-4.5,10.1-10.1,10.1c-5.6,0-10.1-4.6-10.1-10.1c0-5.6,4.5-10.1,10.1-10.1C84.2,46.7,88.7,51.3,88.7,56.8z};
}}
\newcommand\orcidicon[1]{\href{https://orcid.org/#1}{\mbox{\scalerel*{
\begin{tikzpicture}[yscale=-1,transform shape]
\pic{orcidlogo};
\end{tikzpicture}
}{|}}}}

\title{$k$-evolution: a relativistic N-body code for clustering dark energy}
\author[a]{Farbod Hassani~\orcidicon{0000-0003-2640-4460},}
\author[b]{Julian Adamek~\orcidicon{0000-0002-0723-6740},}
\author[a]{Martin Kunz~\orcidicon{0000-0002-3052-7394},}
\author[c]{and Filippo Vernizzi~\orcidicon{0000-0003-3426-2802}}

\affiliation[a]{
Universit\'e de Gen\`eve, D\'epartement de Physique Th\'eorique and CAP,
24 quai Ernest-Ansermet, CH-1211 Gen\`eve 4, Switzerland
}
\affiliation[b]{
School of Physics and Astronomy, Queen Mary University of London, 327 Mile End Road, London E1 4NS, UK}
\affiliation[c]{Institut de physique th\' eorique, Universit\'e  Paris Saclay, 
CEA, CNRS, 91191 Gif-sur-Yvette, France}
\emailAdd{farbod.hassani@unige.ch}
\emailAdd{julian.adamek@qmul.ac.uk}
\emailAdd{martin.kunz@unige.ch}
\emailAdd{filippo.vernizzi@ipht.fr}

\abstract{
We introduce  $k$-evolution, a relativistic $N$-body code based on \gev, which includes clustering dark energy among its cosmological components. 
To describe dark energy, we use the effective field theory approach. In particular, we focus on $k$-essence with a {speed of sound} much smaller than unity but we lay down the basis to extend the code to other dark energy and modified gravity models. We develop the formalism including dark energy non-linearities but, as a first step, we implement the equations in the code after dropping non-linear {self-coupling} in the $k$-essence field. 
In this simplified setup, we compare \kev simulations with those of \class and \gev 1.2, showing the effect of dark matter and gravitational non-linearities on the power spectrum of dark matter, of dark energy and of the gravitational potential. Moreover, 
we compare \kev to Newtonian $N$-body simulations with back-scaled initial conditions and study how dark energy clustering affects massive halos.
}
\graphicspath{Images/}
\begin{document}
\maketitle
\section{Introduction}

The physical reason for the observed acceleration of the Universe \cite{Riess:1998cb} is one of the most important mysteries in cosmology, and arguably generally in fundamental physics. Although cosmology has been revolutionised by the arrival of high quality observations that have allowed to pin down many parameters of the standard model \cite{Aghanim:2018eyx}, the dark sector is still compatible with a cosmological constant and collisionless cold dark matter. This is one motivation for the next generation of large galaxy surveys like \cite{Amendola:2016saw,Santos:2015bsa} 
that will observe billions of galaxies to provide galaxy number counts and weak lensing measurements. 

But much of this data will probe scales that are mildly or strongly non-linear, a regime that is not well modelled even for the Lambda-cold-dark matter ($\Lambda$CDM) standard model \cite{Jalilvand:2019brk}. 
{In this paper we start a systematic study of the effects of dark energy on cosmological structure formation in the non-linear regime. We will do so by implementing dark energy in the relativistic $N$-body simulation code 
\gev \cite{Adamek:2015eda,Adamek:2016zes}. The \gev code works in the weak field limit of General Relativity (GR), which makes it easy to include additional relativistic fields. To model dark energy theories, we will use the Effective Field Theory of dark energy (EFT of DE) approach (see e.g.~\cite{Gubitosi:2012hu} and references below; see also \cite{Frusciante:2017nfr,Cusin:2017mzw} for studies of the non-linear action in the EFT of DE approach and \cite{Cusin:2017wjg} for an application to perturbation theory beyond linear order). The  EFT of DE allows one to describe, in a relativistic setting and with a minimal set of parameters, very general dark energy and modified gravity models based on a single scalar degree of freedom. 
The use of \gev allows to combine both the EFT of DE and the weak field expansion systematically, paving the way for $N$-body simulations of a wide class of dark energy and modified gravity models.

In this first paper we illustrate the approach specifically with the example of $k$-essence  {\cite{ArmendarizPicon:2000dh,ArmendarizPicon:2000ah}. After a brief reminder of the basics of the EFT of DE framework in Sec.~\ref{sec:eft}, we derive the relevant equations (with more details given {in  App.~\ref{sec:conseq}}) and {briefly} discuss their practical implementation in our new code, \kev. {We  refer the reader to   App.~\ref{sec:numimplement} for more details on the numerical implementation of the dark energy equations and to App.~\ref{sec:IC} for a discussion on gauge issues when comparing with linear codes and setting the initial conditions.} 

Then, in Sec.~\ref{sec:results} we present a detailed analysis of the power spectra computed with $k$-evolution and compare the spectra to those obtained with other numerical codes. In Sec.~\ref{sec:snapshots} we examine snapshots of the dark energy simulations computed with $k$-evolution with particular attention to the environment of massive clusters. Finally, we conclude in Sec.~\ref{sec:Conclusion}.}

Due to its complexity, the full non-linear treatment of the dark energy perturbation equations requires a dedicated study that will be addressed in a separate publication. {(See however App.~\ref{sec:smallcslimit} where we  study the evolution of non-linear perturbations in the limit of small speed of sound and in matter domination.)} 
In this paper we limit the simulations to the linear $k$-essence equations, although these equations are coupled to the non-linear clustering of the dark matter, which can in turn lead to non-linear clustering of the dark energy. This approach will be called \kev in the following. We compare it to the fully linear treatment implemented in version 1.2 of {\em gevolution}\footnote{\url{https://github.com/gevolution-code/gevolution-1.2.git}} that uses a realisation of the linear fluid transfer functions from \class \cite{Blas:2011rf} in the $w$-$c_s^2$-parametrisation and is presented here for the first time, as well as to a standard $N$-body simulation where only the background evolution is changed according to the equation of state $w$, without allowing for perturbations in the dark energy.  


\section{The EFT of $k$-essence\label{sec:eft} }

In this section we introduce the effective field theory description of dark energy and we  derive the relevant equations for its implementation in \gev. We assume that matter -- the dark matter and the Standard Model particles -- is minimally coupled to the gravitational metric $g_{\mu \nu}$. Moreover, as explained in the introduction, we consider theories with a preferred time-slicing induced by the evolution of a scalar field and
 we  focus on operators describing general scalar-field Lagrangians that can be constructed out of the field value $\phi$ and  its first derivatives contracted with  $g_{\mu \nu}$, i.e., $X \doteq g^{\mu \nu} \partial_\mu \phi \partial_\nu \phi$. {In the covariant language, 
 the action describing this class of theories is 
 \begin{equation}
S_{\rm DE}= \int  d^4 x \sqrt{-g} P (X, \phi)  \;,\label{eq:kessence_action}
 \end{equation}
which is also known as   
$k$-essence \cite{ArmendarizPicon:2000dh,ArmendarizPicon:2000ah}.}

\subsection{The  action and the homogeneous equations}

To describe the dark energy fluctuations we adopt the EFT of DE description {\cite{Creminelli:2008wc,Gubitosi:2012hu,Bloomfield:2012ff,Gleyzes:2013ooa,Bloomfield:2013efa} (see also \cite{Piazza:2013coa,Tsujikawa:2014mba,Gleyzes:2014rba,Frusciante:2019xia} for reviews; for other effective relativistic approaches, see for instance \cite{Baker:2011jy,Baker:2012zs,Battye:2012eu,Bellini:2014fua,Lagos:2016wyv})}, which
is particularly convenient {for studying} fluctuations around cosmological FRW solutions with a preferred slicing induced by the time-dependent background scalar field. 
In the  unitary gauge, where the time coincides with uniform-field hypersurfaces, the EFT action expanded around a spatially flat background reads \cite{Cheung:2007st,Creminelli:2008wc}  
\be
\label{EFTaction}
S=\int d^4 x \sqrt{-g} \left [ \frac{\MP^2}{2} R -\Lambda (t) -c(t) g^{00} +\frac{M_2^4(t)}{2} \left (\delta g^{00}  \right )^2    + \ldots  \right ] \;,
\ee
where $R$ is the four-dimensional Ricci scalar, $\Lambda(t)$, $c(t)$, and $M_2^4(t)$ are time-dependent functions and $\delta g^{00}$ is the perturbation of $g^{00}$ around its homogeneous value. 
The {ellipsis stands} for terms that are of higher order in the fluctuations $\delta g^{00}$. These terms can be  ignored because they are negligible in the weak-field expansion adopted by \gev (see e.g.~Refs.~\cite{Cusin:2017mzw} for details).  We will come back to this point at the beginning of  Sec.~\ref{sec:perts}.

The functions $\Lambda(t)$ and $c(t)$ are not independent; they can be expressed in terms of the background expansion and  matter quantities through the homogenous Friedmann equations. Varying the action with respect to the homogenous lapse $N(t)$ and the scale factor $a(t)$, appearing in the homogenous metric as $ds^2 = -N^2(t) dt^2 +a^2(t) d\vec x^2 $, we obtain 
\begin{align}
\frac{\dot a^2}{a^2} \doteq H^2 & = \frac{1}{3 \MP^2} \left( \rho_{\rm m} + c + \Lambda   \right)\;, \label{frie1} \\
\frac{\ddot a}{a} = \dot H + H^2 & = - \frac{1}{6 \MP^2} \left( \rho_{\rm m} + 3 p_{\rm m} + 4 c - 2 \Lambda    \right) \label{frie2} \;,
\end{align}
where $\rho_{\rm m}$ and $p_{\rm m}$ are respectively the homogeneous matter energy density and pressure. 

The stress-energy tensor of dark energy can be computed from the above action {as}
\be
\label{SEtensor}
T_{\mu \nu} = - \frac{2}{\sqrt{-g}} \frac{\delta S_{\rm DE}}{\delta g^{\mu \nu}} \;,
\ee
where $S_{\rm DE}$ is the action  without the Einstein-Hilbert term.
In unitary gauge one finds
\be
T_{\mu \nu} = - \left[  \Lambda + c g^{00} - \frac{M_2^4}{2} (\delta g^{00})^2 \right] g_{\mu \nu} + 2 (c  - M_2^4 \delta g^{00} ) \delta_\mu^0 \delta_\nu^0 \;.
\ee
The homogeneous part of the stress-energy tensor is obtained by taking {$g^{00}=-1$ and} $\delta g^{00}=0$ in the above expression. Rewriting its components in terms of the homogeneous energy density $\rho_{\rm DE}(t)$ and pressure $p_{\rm DE}(t)$ of the dark energy, using $T_{00} = \rho_{\rm DE}(t)$ and $T_{ij} = \delta_{ij} p_{\rm DE}(t) $, we obtain
\be
\label{rhop}
c(t) = \frac12 \left[ \rho_{\rm DE}(t)+p_{\rm DE}(t) \right] \;, \qquad \Lambda(t) = \frac12 \left[\rho_{\rm DE}(t) - p_{\rm DE}(t) \right] \;,
\ee
which is consistent with the Friedmann equations above. Moreover, using these expressions, taking the derivative with respect to time of Eq.~\eqref{frie1} and using Eq.~\eqref{frie2} we can also check that the Friedmann equations are consistent with the homogeneous continuity equation {for dark energy, i.e.,} 
\be
\label{SEcont}
\dot \rho_{\rm DE} + 3 H (\rho_{\rm DE} + p_{\rm DE}) =0 \;,
\ee 
as expected. {From Eq.~\eqref{rhop}, this implies that $c$ and $\Lambda$ satisfy $\dot c + \dot \Lambda + 6H c=0$.}

\subsection{St\"uckelberg trick to conformal time}

To study the cosmological perturbations in this setup, it is convenient to adopt a gauge where the perturbations of the scalar field are explicit.
We can restore diffeomorphism invariance of the action by the  St\"uckelberg trick \cite{Cheung:2007st,Gleyzes:2013ooa}, i.e.,~by performing a time-diffeomorphism 
\be
\label{time_diff}
t \to \tilde t = t+ \xi^0 (t , \vec x) \;, \qquad \vec x \to \vec {\tilde x} = \vec x  \; ,
\ee
and promoting the parameter $\xi^0$ to a field. 
In the following, {however,} instead of using cosmic time $t$ we will {present the evolution equations with the conformal time $\eta$,  which is related to $t$ by $\eta \doteq \int dt/a(t)$. 
For this reason, {instead of defining the Goldstone boson of  broken time diffeomorphisms $\pi$ as  the parameter of the time diffeomorphism $\xi^0$ \cite{Cheung:2007st}, we will adopt a definition adapted to the conformal time $\eta$. In particular, we will define $\pi$ as} $\xi^0$ divided by the scale factor, i.e.,
\be
\pi(t,\vec x) \doteq \xi^0 (t,\vec x) /a (t) \;.
\ee 

{We can now compute how the quantities in the action \eqref{EFTaction} change under the time-transformation above, paying attention to expressing the cosmic time quantities in terms of conformal time.}
The Ricci scalar does not change under the transformation \eqref{time_diff},  while $g^{00}$ transforms as
\be
\begin{split}
g^{00} (t, \vec x)  \to \tilde g^{00} ( \tilde t (\eta,\vec x), \vec x) &=  \frac{\partial \tilde t (\eta,\vec x) }{\partial x^\mu}  \frac{\partial \tilde t (\eta,\vec x)}{\partial x^\nu}  g^{\mu \nu} (  t (\eta), \vec x)  \\
&= \frac{\partial \left[ t (\eta) + a(\eta) \pi(\eta,\vec x) \right]}{\partial x^\mu}  \frac{\partial \left[ t (\eta) + a(\eta) \pi(\eta,\vec x) \right] }{\partial x^\nu} g^{\mu \nu} (\eta, \vec x) \;, 
\end{split}
\ee
which gives
\be
\label{stukg00}
g^{00} (t, \vec x)  \to \tilde g^{00} (  t,  \vec x) =   a^2  \big[ (1+ \HH \pi)^2 g^{00} + 2 (1+\HH \pi) g^{0 \mu} \partial_\mu \pi + g^{\mu \nu} \partial_\mu \pi \partial_\nu \pi  \big]\;,
\ee
where the untilded metric on the right-hand side is the one  in conformal time.

Any function of time, instead, transforms as
\be
\begin{split}
\label{fchange}
f(t) \to  \tilde f \left( \tilde t (\eta,\vec x) \right)& = f \left(t (\eta) + a (\eta) \pi(\eta,\vec x) \right) \\
&= f(\eta) + f'(\eta)  \pi(\eta,\vec x) + \frac12 \left[f'' (\eta) - \HH (\eta) f'(\eta) \right] \left[\pi (\eta,\vec x) \right]^2
 + \ldots  \;,
\end{split}
\ee
where $\HH \doteq a'/a $ is the conformal Hubble rate.
Applying these transformations to the  action \eqref{EFTaction}, we can derive the fully covariant action,
\be
\label{EFTaction_cov}
\begin{split}
S=\int d^4 x \sqrt{-g} \bigg\{ & \frac{\MP^2}{2} R -  \Lambda \left[ t(\eta)+ a(\eta) \pi \right] - c \left[ t(\eta)+ a(\eta) \pi \right]  a^2(\eta)  \big[ (1+ \HH \pi)^2 g^{00} \\
&+ 2 (1+\HH \pi) g^{0 \mu} \partial_\mu \pi + g^{\mu \nu} \partial_\mu \pi \partial_\nu \pi  \big]  +\frac{M_2^4 \left[ t(\eta)+ a(\eta) \pi \right]}{2}  \\
& \times a^4 (\eta)  \big[ (1+ \HH \pi)^2 g^{00} + 2 (1+\HH \pi) g^{0 \mu} \partial_\mu \pi + g^{\mu \nu} \partial_\mu \pi \partial_\nu \pi  - \bar g^{00}\big]^2      \bigg\} \;,
\end{split}
\ee
where we have used that the Ricci scalar $R$ does not transform, and in the last term we have introduced $\bar g^{00}$, the background value of $g^{00}$, i.e.,~$\bar g^{00} = -1/a^2$.

Finally, using the definition \eqref{SEtensor}, we can write the  expression of the stress-energy tensor, which reads
\be
\label{SEfull}
\begin{split}
T_{\mu \nu} = & \ - \Big\{ \Lambda \left( t+ a  \pi \right)  + c \left( t+ a \pi \right)  a^2  \big[ (1+ \HH \pi)^2 g^{00} + 2 (1+\HH \pi) g^{0 \rho} \partial_\rho \pi + g^{\rho  \sigma} \partial_\rho \pi \partial_\sigma \pi  \big]  \\
&- \frac{M_2^4 \left( t+ a \pi \right)}{2}   a^4 \big[(1+ \HH \pi)^2 g^{00} + 2 (1+\HH \pi) g^{0 \rho} \partial_\rho \pi + g^{\rho  \sigma} \partial_\rho \pi \partial_\sigma \pi - \bar g^{00} \big]^2  \Big\}g_{\mu \nu}  \\ 
& + 2 \big[(1+ \HH \pi)^2 \delta^0_\mu \delta^0_\nu + 2 (1+\HH \pi) \delta^0_{(\mu} \partial_{\nu)} \pi +  \partial_\mu \pi \partial_\nu \pi  \big]  \Big\{   c \left( t+ a  \pi \right) a^2 \\
&- M_2^4 \left( t+ a  \pi \right) a^4   \big[ (1+ \HH \pi)^2 g^{00} + 2 (1+\HH \pi) g^{0 \rho} \partial_\rho \pi + g^{\rho  \sigma} \partial_\rho \pi \partial_\sigma \pi  - \bar g^{00}\big] \Big\}   \;.
\end{split}
\ee
This expression is fully nonlinear and can be expanded at any given order in perturbations. In deriving it, we only assumed that higher powers of $\delta g^{00}$ in the action \eqref{EFTaction} are negligible in the weak field limit, which {we justify below.}  

\subsection{Perturbations }
\label{sec:perts}
To study the perturbations, we will use the Poisson gauge, where the metric reads
\be
\label{Poissonmetric}
ds^2 = a^2(\eta) \left[  - e^{2 \Psi} d\eta^2-2 B_i dx^i dt + ( e^{-2 \Phi} \delta_{ij} + {h}_{ij} ) dx^i dx^j \right] \;,
\ee
where $\delta^{ij} \partial_j B_i =0$ and ${\delta^{ij} h_{ij}} = 0 = \delta^{ij}\partial_i h_{jk}$.

In \gev it is assumed that the metric perturbations remain small at the scales of interest. This is implemented by defining a small parameter $\epsilon$, such that $\Phi $,  $\Psi$, $B_i$ and ${h}_{ij}$ are at most of order ${\cal O}(\epsilon)$. For non-relativistic sources, time derivatives are of  order Hubble and do not change the order of a term in the expansion. 
Instead, each spatial derivative lowers the order of a given term by one half, so that for instance $\partial_i \Phi \sim  {\cal O} (\epsilon^{1/2})$ and $\nabla^2 \Phi \sim  {\cal O} (\epsilon^{0})$, where $\nabla^2 \doteq \delta^{ij} \partial_i \partial_j$ defines the Laplacian. 
{We refer the reader to Refs.~\cite{Adamek:2017uiq, Adamek:2016zes} for details.}

The Einstein-Hilbert term in the action \eqref{EFTaction} contains at least two spatial derivatives {of the metric} so that the order of this term is $n-1$, where $n$ is the order of the expansion in metric perturbations. For instance, in \gev one expands the Einstein tensor up to second order in the metric perturbations, which for the terms containing two spatial derivatives corresponds to going at most at order ${\cal O}(\epsilon)$ in the equations of motion. The Einstein tensor up to this order can be obtained by varying the Einstein-Hilbert term expanded up to order ${\cal O}(\epsilon^2)$. To be coherent with this scheme, we have  to keep all the terms in the action that contribute at most to ${\cal O}(\epsilon^2)$.

To evaluate the order of an operator in the EFT of DE action, we need to look at the scalar field perturbation $\pi$.
On large scales, linear cosmological perturbation theory is recovered. In this case $\pi$ is of the same  order as  the metric perturbations. 
In particular, using the scaling above we have
\be
\pi \sim {\cal O} (\epsilon) \;, \qquad \partial_i \pi \sim {\cal O} (\epsilon^{1/2}) \;, \qquad \nabla^2 \pi \sim {\cal O} (\epsilon^0) \;.
\ee
For instance, this means that we need to expand the operator $\Lambda$ up to second order in $\pi$ using Eq.~\eqref{fchange}.
Moreover, by Eq.~\eqref{stukg00} $\delta g^{00}$ is at least of order ${\cal O}(\epsilon)$, which implies that any operators of order higher that $(\delta g^{00})^2$ can be neglected, which justifies truncating the action  \eqref{EFTaction} at this order.

We can now discuss the field equations.
Variation of the action with respect to $\pi$ gives the evolution equation for the perturbation of the scalar field.  For later purposes, it is convenient to introduce the variable 
\be
\zetaf \doteq \pi ' + \HH \pi - \Psi \;, \label{pi_eq}
\ee
and express the first and second time-derivative of $\pi$ in terms of $\zetaf$ and $\zetaf'$. Using the conservation equation \eqref{SEcont}, which in terms of the conformal time reads 
\be
\rho'=-3\HH (\rho+p) \;, 
\ee
by varying the action we  obtain the following system of coupled equations:
\begin{align}
 \pi '&  = \zetaf - \HH \pi + \Psi \;, \label{zeta_eq1} \\
 \zetaf' & =  (3 c_a^2 + s) \HH \zetaf -  3 c_s^2 \left(\HH^2 \pi - \HH \Psi - \HH' \pi - \Phi' \right) +c_s^2 \nabla^2 \pi  \nonumber\\
& - \vec  \nabla \left[2 (c_s^2-1)  \zetaf +c_s^2    \Phi-    \Psi \right] \cdot \vec  \nabla \pi - \left[ (c_s^2-1)\zetaf    + c_s^2 \Phi - c_s^2  \Psi \right] \nabla^2 \pi  \nonumber  \\
& -  \frac{\HH}{2} \left[ (2+ 3 c_a^2 +c_s^2 + s) (\vec \nabla \pi)^2 +  6 c_s^2 (1+ c_a^2) \pi  \nabla^2 \pi  \right]     +\frac{c_s^2 -1 }{2} \partial_i \left( \partial_i \pi (\vec \nabla \pi)^2 \right) \;, \label{zeta_eq2}
\end{align}
where  we have introduced the speed of propagation {squared} of {dark energy} fluctuations, $c_s^2$, its rate of change, $s$, and the adiabatic {speed of sound squared} (which generally differs from the speed of propagation) $c_a^2$. These are respectively defined as\footnote{ The covariant $k$-essence action,  Eq.~(\ref{eq:kessence_action}), implies that \cite{Bonvin:2006vc} $w = \frac{P}{2 X P_{,X}-P}$ and $c_{s}^{2}=\frac{P_{,X}}{2 X P_{, XX}+P_{,X}}$, where we have denoted the symbol of partial derivation by a comma.}
\be
c_s^2 \doteq \frac{c}{c+2 M_2^4} \;, \qquad s\doteq \frac{(c_s^2)'}{c_s^2 \HH} \;, \qquad c_a^2 \doteq \frac{p'}{\rho'} = \frac{\Lambda' - c'}{c' + \Lambda'}\;,
\ee
where for the last equality we have used Eq.~\eqref{rhop}. Notice that $c_a^2$ can be related to the time derivative of the equation of state $w \doteq p/\rho$ by $w'=-3\HH (1+w) (c_a^2 - w)$ so that $w$ and $c_s^2$ completely characterize the model. 

Let us pause to comment {on} these equations.  First, all the terms are of order ${\cal O}(\epsilon)$, {with the exception of $\nabla^2 \pi$. This term is ${\cal O}(1)$ in our perturbative scheme. But this term generates the pressure support within the sound-horizon of the scalar field, and leads to wave-like behaviour, not a growth of perturbations. For this reason it does not change the order of $\pi$ which is an ${\cal O}(\epsilon)$ quantity. The other terms are all small; as an example we can consider the last term. It} involves three fields $\pi$ (so it is of order 3 in the standard perturbation-theory expansion), and since it contains four spatial derivatives its order is $3-4/2=1$. Second, the equations are at most of order three in the perturbations, which is a consequence of the fact that we are considering only theories with at most one derivative per field in the action so that one pays at least an ${\cal O}(\epsilon^{1/2})$ for each new order in the perturbations.
Third, the limit of $c_s^2 \to 1$, obtained by sending $M_2^4 \to 0$, is well defined. In this case the last cubic term vanishes but this is to be expected because it can only come from the operator $M_2^4$ in the action \eqref{EFTaction}. Also the limit $c_s^2 \to 0$ (obtained for $M_2^4 \gg c$) is well defined; we will come back to it {at} the end of the section. 

Let us turn now to the stress-energy tensor of dark energy.
Expanding Eq.~\eqref{SEfull}  using the Poisson metric \eqref{Poissonmetric} one obtains
\be
\label{SEcomponents}
\begin{split}
 T_0^0 &= - \rho  + \frac{\rho + p}{c_s^2}  \bigg[ 3c_s^2 \HH \pi -\zetaf - \frac{2c_s^2 - 1}{2} (\vec{\nabla} \pi)^2   \bigg ] \;, \\
 T^0_i & = - (\rho+p) \bigg[1- \frac{1}{c_s^2}   \, \Big( 3 c_s^2 (1+w) \mathcal{H} \pi -\zetaf + c_s^2 \Psi \Big) + \frac{c_s^2 - 1}{2 c_s^2}   (\vec{\nabla} \pi)^2  \bigg ] \partial _i \pi  \; ,\\ 
T_{j}^{i}&  =  p \delta_{j}^{i}  - (\rho+p)  \bigg[ 3 c_a^2  \HH \pi -\zetaf +  \frac12 (\vec{\nabla} \pi)^2   \bigg] \delta_{j}^{i}  + (\rho+p) \delta^{i k} \partial_k \pi \partial_j \pi  \;,
\end{split}
\ee
where  we have used the homogeneous continuity equation and we have expanded $T_0^0$ and $T_i^j$ up to order ${\cal O}(\epsilon)$ and $T_0^i$ up to order ${\cal O}(\epsilon^{3/2})$.

The latter is expanded to higher order than $T_0^0$ and $T_i^j$ because  its divergence, which is ${\cal O} (\epsilon)$, appears as  the source in the continuity equation. 
As shown in Appendix~\ref{sec:conseq}, Eqs.~\eqref{zeta_eq1} and \eqref{zeta_eq2} are equivalent to the conservation equation $\nabla_\mu T^\mu_\nu=0$ of the stress-energy tensor above. 
Note that in the limit $c_s^2 \to 0$ the components $T_0^0$ and $T^0_i$ seem to blow up due to the $1/c_s^2$ term. 
However, one can show that the brackets on the right-hand side of these expressions vanish at leading order in $c_s^2$, so that  the stress-energy tensor remains finite.
We discuss this in more  detail in the case of matter domination in App.~\ref{sec:smallcslimit}. 

We also note that when linearized, the stress-energy tensor is purely scalar. The higher-order terms do not preserve this property, but the resulting vector and tensor type contributions will be small. For this reason we do not expect scalar dark energy to lead to significantly larger vector and tensor perturbations than $\Lambda$CDM.
The dark energy stress-energy tensor must be inserted in the Einstein equations (obtained from the variation of Eq.~\eqref{EFTaction} with respect to the metric), {together with} the stress-energy tensors of the other species. 
The Einstein equations in the weak field approximation are
\be
 {(1+2 \Phi) \nabla^2 \Phi-3 \mathcal{H} \Phi^{\prime}-3 \mathcal{H}^{2}(\Phi-\chi)-\frac{1}{2} \delta^{i j} \partial_i \Phi \partial_j \Phi=-4 \pi G a^{2} \delta T_{0}^{0}} \;,
\ee
\be
\nabla^4 \chi-\left(3 \delta^{i k} \delta^{j l} \frac{\partial^{2}}{\partial x^{k} \partial x^{l}}-\delta^{i j} \nabla^2\right) \Phi_{, i} \Phi_{, j}=4 \pi G a^{2}\left(3 \delta^{i k} \frac{\partial^{2}}{\partial x^{j} \partial x^{k}}-\delta_{j}^{i} \nabla^2 \right) T_{i}^{j} \;,
\ee
where $\nabla^4 \doteq \delta^{ij}  \delta^{kl} \partial_i \partial_j  \partial_k \partial_l$ and the stress tensor $T_{\mu}^{\nu}$ includes the relevant species including matter and dark energy, $\chi \doteq \Phi-\Psi,$ and the transverse projection tensor is defined as,
\be
P_{i j} \doteq \frac{\partial^{2}}{\partial x^{i} \partial x^{j}}-\delta_{i j} \nabla^2 \;.
\ee
Here we do not discuss the equations for vector and tensor perturbations, as we are not going to study them in this paper.

\subsection{Implementation}
\label{Sec1}
In this work, we remove the non-linear terms in the {$\pi$ evolution} equations and stress-energy tensor. Due to their complexity, we are going to study the non-linear self-interaction of dark energy in detail in a separate work. 
It is interesting to note that although we have removed the {$\pi$} non-linear self-interaction, the energy density of the scalar field nonetheless becomes non-linear as it is sourced by matter going non-linear.
{For the sake of simplicity,} we also assume that both $w$ and $c_s^2$ are constant, which implies 
\be
s=0 \;, \qquad c_a^2 = w \;.
\ee
Theoretically, this is not well motivated but it would not be difficult to include the time-evolution of $w$ and $c_s^2$. However, since there are no especially well-motivated models in any case, we prefer to consider here only the simplest scenario.

When we neglect the non-linear terms, the {$\pi$ evolution} equations \eqref{zeta_eq1} and \eqref{zeta_eq2}  read
\begin{align}
 \pi '&  = \zetaf - \HH \pi + \Psi \;, \label{zeta_eq1_lin} \\
 \zetaf' & =  3w \HH \zetaf -  3 c_s^2 \left(\HH^2 \pi - \HH \Psi - \HH' \pi - \Phi' \right) +c_s^2 \nabla^2 \pi\;, \label{zeta_eq2_lin} 
\end{align}
and the linear stress tensor becomes
\be
\label{SEcomponents_lin}
\begin{split}
 T_0^0 &= - \rho  + \frac{\rho + p}{c_s^2}  \Big( 3c_s^2 \HH \pi -\zetaf   \Big) \;, \\
 T^0_i & = - (\rho+p) \partial _i \pi  \; ,\\ 
T_{j}^{i}&  =  p \delta_{j}^{i}  - (\rho+p)  \Big( 3 c_a^2  \HH \pi -\zetaf  \Big) \delta_{j}^{i}  \;.
\end{split}
\ee
A detailed description of the numerical implementation can be found in Appendix\ \ref{sec:numimplement}.

\section{Numerical results for power spectra \label{sec:results}}

In this section we compare the power spectra from \kev with the linear perturbation solutions from \class~{\cite{Lesgourgues:2011re}} and with {the power spectra computed with} \gev 1.2 using the \class  interface to include {dark energy}. For both cases we consider two different {speeds of sound}: $c_s^2 = 10^{-7}$ and $c_s^2 = 10^{-4}$.
We also test the effects that arise when trying to simulate a different expansion history in a Newtonian simulation without including a dark energy fluid at all.

We always combine two simulations with sizes $L=9000 \; \text{Mpc}/h$ and $L=1280  \; \text{Mpc}/h$, both {with} a grid of size $N_\text{grid}=3840^3$.  All the results in this section have been obtained using the cosmological parameters shown in {Tab.~}\ref{table:1}.
\renewcommand{\arraystretch}{2.2}
\begin{table}
\begin{center}
\begin{adjustbox}{max width=\textwidth}
\begin{tabular}{|c|c |c |c | c| c| c| c|c|    } 
\hline
$h$ & $n_s$ &  $A_s$ &  $\Omega_{\rm b} h^2$   & $\Omega_{\rm CDM} h^2$  &   $\Omega_{\rm DE} $	&	T$_{\rm CMB}[K] $    &      	$N_{\nu}$  & $w$	   \\
\hline
 0.67556 & 0.9619                    &  $ 2.215 \times 10^{-9}$  &   0.02203  &    0.12038 &  			 0.68786            	&       2.7255		&	  3.046	&  -0.9    \\
\hline
  \end{tabular}
\end{adjustbox}
\end{center}
\caption{{Values of the cosmological parameters used in this paper. In particular, $n_s$ and $A_s$ are respectively the  spectral index and amplitude  of the primordial scalar fluctuations; $\Omega_{\rm b}$, $\Omega_{\rm CDM}$ and $\Omega_{\rm DE}$ are  the critical densities, respectively of  baryons, CDM and dark energy; {$h \doteq H_0 / (100 \, \text{Km}\, s^{-1} \text{Mpc}^{-1})$ is the  reduced Hubble constant;} $N_{\nu}$ is the Standard Model effective number of neutrino species while $w$ is the equation of state of dark energy. We also consider pivot wavenumber $k_p = 0.05$ Mpc$^{-1}$.}}
\label{table:1}
\end{table}

In {Fig.~}\ref{fig:codes} we {illustrate the conceptual} difference between the three codes we use in this section: In \kev matter and {gravitational} potentials are treated non-linearly\footnote{With non-linear gravitational potentials we do not mean that they become large, 
but that they are different from the linear predictions {e}specially {for} large wave numbers. {However, they still remain} small and respect the weak-field approximation.} and {act as a source for the linearized dark energy equations so that the dark energy contains non-linear contributions as well}. This is consistently taken into account when} dark energy density sources the gravitational potential. In \gev the dark energy density is approximated by its linear solution {computed with \class} and is therefore not sourced by the non-linearities of matter. {However, the gravitational} potentials are sourced by {this} linear dark energy density{, and matter evolves accordingly}. Finally, in \class all the fields are linearized and all species source each other. 
\begin{figure} [H]
\centering
 \includegraphics[scale=0.3]{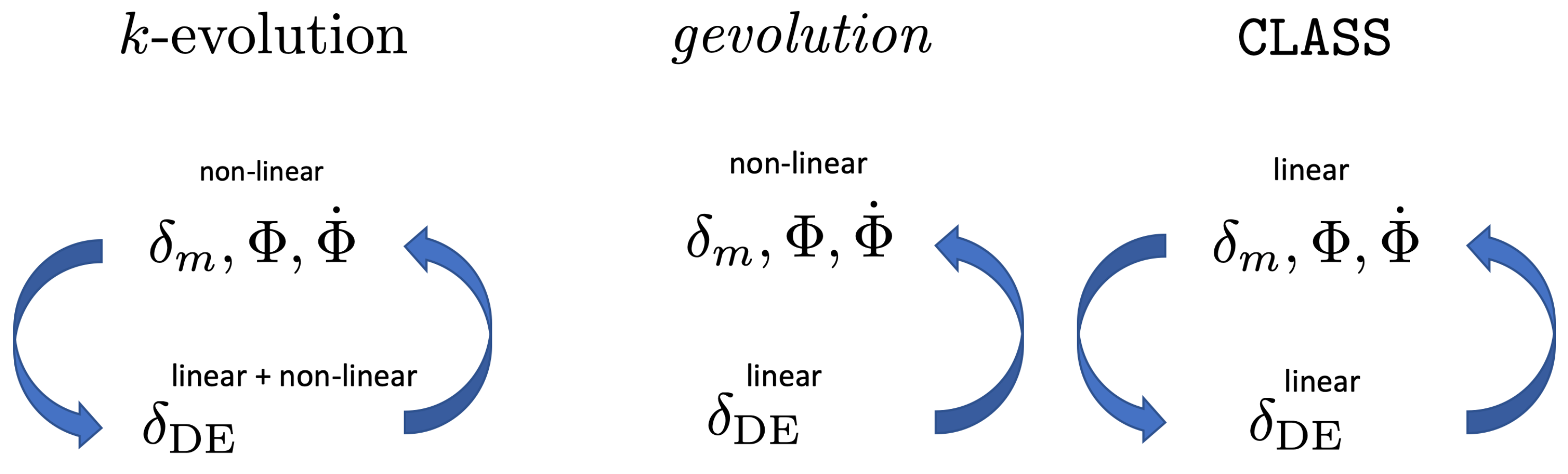} 
 \caption{{ Schematic description of the three different codes used in this work, where} 
 the blue arrows show 
 if one component sources the other.
 {On the left, the two blue arrows {going} from non-linear matter and {potentials} to $\delta_{ \rm DE}$, {and {\em vice-versa}, show} that in \kev all the components interact and source each other. Although we have used the density {of linearized dark energy,} it{s solution} becomes non-linear since it is sourced by other species.} In \gev the matter and potentials become non-linear and are sourced by linear dark energy density, while $\delta_{\rm DE}$ is {computed with \class}. In \class all the components are linear and interact with each other.}
 \label{fig:codes}
 \end{figure}

\subsection{\kev versus \class}

We start by comparing \kev with the linear Boltzmann code \class. In \class code, one can extend the {matter} power spectrum beyond the linear regime by the use of Halofit \cite{Takahashi:2012em}. However, one should remember that Halofit is calibrated to simulations without clustering dark energy. In the following we will use \class both with and without the use of Halofit. The power spectrum of a given quantity $X$ is defined by
\begin{equation}
\langle \hat X(\vec k) \hat  X(\vec k') \rangle = (2 \pi)^3 \delta (\vec k + \vec k') P_X(k)\;,
\end{equation}
where $\hat  X(k)$ is the Fourier transform of $X$.
The dimensionless power spectrum is defined by 
\begin{equation}
    \Delta_X(k) \doteq \frac{k^3}{2 \pi^2} P_X(k) \;.
\end{equation}

\begin{figure}%
      \centering
       \includegraphics[width=0.95\textwidth]{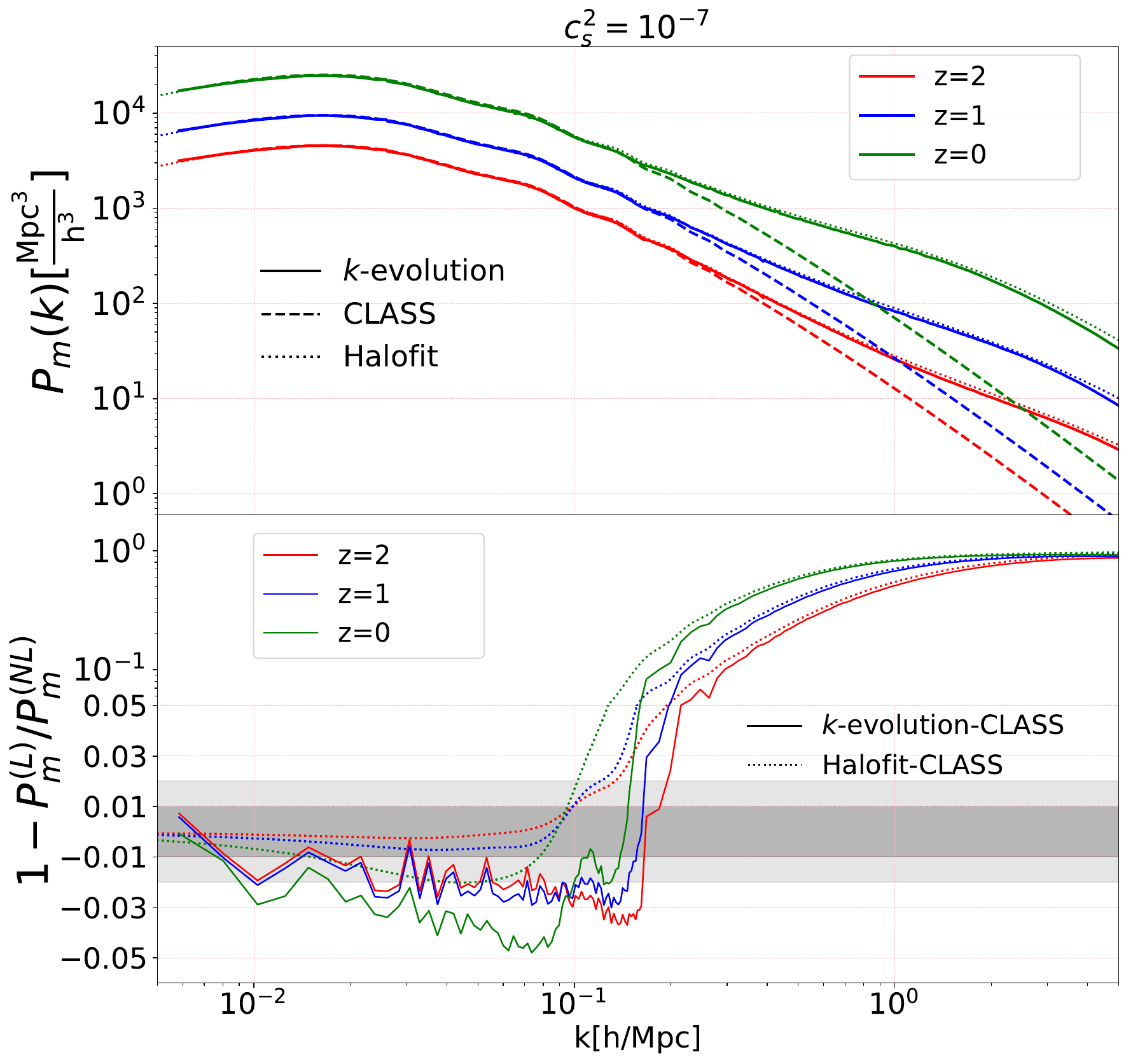}
     \caption{The top panel shows the matter power spectra from 
     \kev and \class at different redshifts for $c_s^2=10^{-7}$. The bottom panel shows the relative difference between the matter power spectra of \class-Halofit and those of \class-\kev. The relative difference {increases in the quasi-linear regime} and {for $z=0$} reaches $\sim -5\%$ at its peak for \class-\kev and $\sim -2 \%$ for \class-Halofit, while at high wavenumbers the non-linearity dominates. 
     {Notice that the vertical axis of the bottom panel is logarithmic above $0.05$ and linear below.} }
  \label{fig:dm_class_kev}
 \end{figure}

In Fig.~\ref{fig:dm_class_kev} we show the matter power spectrum. The onset of non-linear structure formation is clearly visible on scales $k>0.1 h$/Mpc, where the relative difference between the linear and non-linear power spectra changes sign. On very large scales, \kev agrees with \class at the percent-level for all redshifts, but at intermediate scales {and} at low redshifts the difference increases to about 5\%. 
We will see in Sec.~\ref{sec:jev}, where we compare our results from \kev with  \gev,  that the agreement there at low redshifts is much better, which means that the relative difference here comes from the effect of non-linearities at quasi-linear scales. {Here we only plot the results for {$c_s^2 = 10^{-7}$}} as we will show in {Sec.~}\ref{sec:jev} that the effect of dark energy clustering on the matter power spectrum is negligible {so that this plot would look the same for other {values of $c_s^2$}.}

The power spectrum from \class extended  beyond the linear regime using Halofit exhibits similar features as the one from \kev, particularly in the quasi-linear regime: the spectra of both \kev\ and Halofit are slightly suppressed relative to the linear power spectrum due to non-linearities\footnote{This comes from the fact {that} in the one-loop contribution to the matter power spectrum in standard perturbation theory, i.e.,~$P^{\rm SPT}_{22} + P^{\rm SPT}_{13}$, the term $P^{\rm SPT}_{13}$ is always negative and is the dominant term at large and quasi-linear scales, up to $\sim 0.1 h/\rm{Mpc}$, while at smaller scales $P^{\rm SPT}_{22}$ becomes the dominant term \cite{Jalilvand:2019brk}.}. For Halofit the differences reach $\sim- 2\%$, {not quite in agreement with} the {$k$-evolution} result, which may be due to the fact that Halofit is not calibrated for such models.

\begin{figure}%
    \centering
       \includegraphics[width=0.95\textwidth]{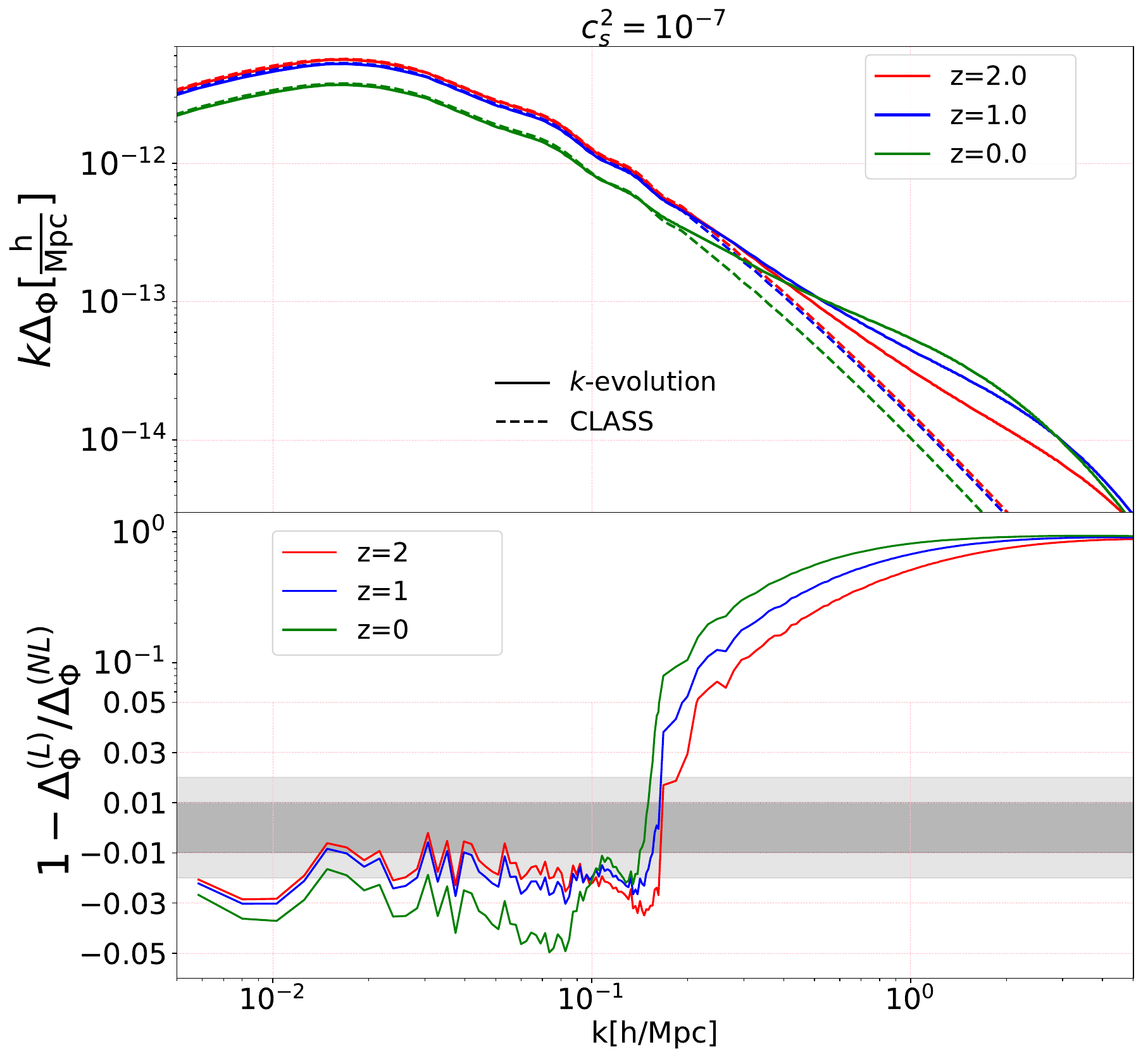}
     \caption{{Comparison of the  power spectra of the gravitational potential $\Phi$ computed with  \kev (solid lines) and \class (dashed lines), for $c_s^2=10^{-7}$. At the top, we show the power spectra at different redshifts and at the bottom their relative difference. On intermediate scales, the power spectra computed with \kev are suppressed compared to those of \class, while the situation is reversed in the non-linear regime, for $k > 0.1 h/$Mps. }}
  \label{fig:pot_class_kev}
 \end{figure}
The situation is similar for the gravitational potential shown in Fig.~\ref{fig:pot_class_kev}, where again we observe the onset of non-linearity at $k>0.1 h$/Mpc, as well as a scale dependent difference at intermediate scales where the linear power spectrum is larger than the non-linear one.
Again, the $\Phi$ power spectra for the two speeds of sound are only slightly different, so that we only show the results for {$c_s^2 = 10^{-7}$} and we will study the effect of  the dark energy speed of sound on the gravitational potential in detail by comparing the potential power spectrum from \kev with \gev in Sec.~\ref{sec:jev} .

\begin{figure}%
    \centering
       \includegraphics[width=1.0\textwidth]{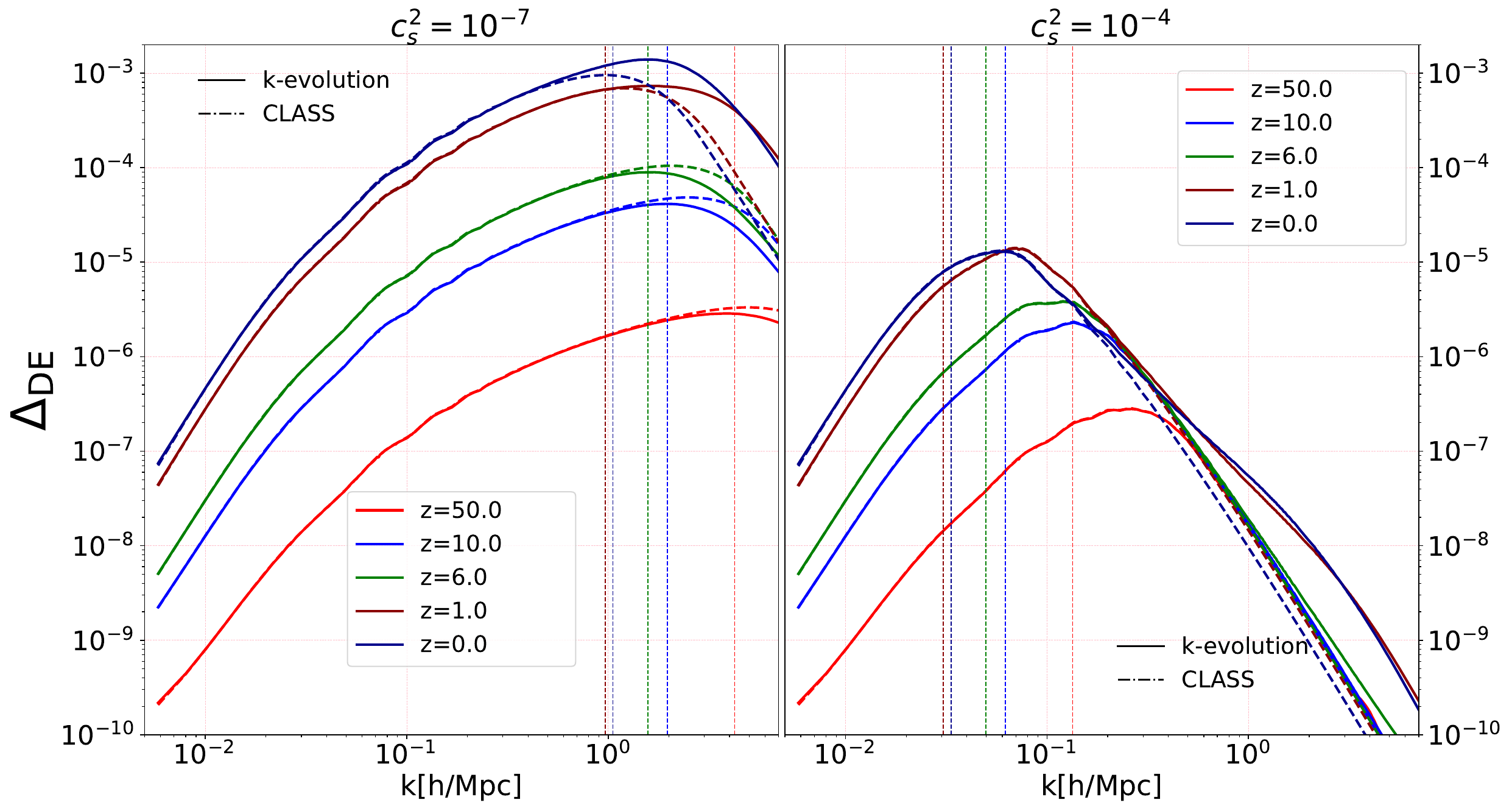}
\caption{{Comparison of the dark energy density power spectra from \kev (solid lines) and from \class (dashed lines) at different redshifts, for two different speeds of sound, $c_s^2=10^{-7}$ (left-panel) and $c_s^2=10^{-4}$ (right-panel). The vertical dashed lines show the value of the dark energy sound-horizon at each redshift, using the same color as the corresponding power spectrum. The turn-around in the power spectra takes place inside the sound-horizon. Its exact position is affected by dark matter non-linearities, as one can see comparing the case $c_s^2=10^{-4}$, where the turn-around happens on linear scales, and the case $c_s^2=10^{-7}$, where the turn-around takes place in the non-linear regime. Notice also that, as for the matter power spectrum, in the linear and quasi-linear regime at $z=0$, the non-linear dark energy power spectrum is smaller than the linear one.} }
  \label{fig:deltakess_class_kev}
\end{figure} 
The dark energy density power spectra  at different redshifts, for $c_s^2=10^{-4}$ and $c_s^2=10^{-7}$,  are shown in Fig.~\ref{fig:deltakess_class_kev}. The vertical lines indicate the sound-horizon of dark energy, which roughly corresponds to the peak of the linear dark energy density power spectrum, since on scales smaller than the sound-horizon the perturbations decay while on scales larger than {the} sound-horizon the perturbations grow. As a result, the density power spectrum of  dark energy  for $c_s^2=10^{-7}$ is much higher than the same quantity for $c_s^2=10^{-4}$: the sound-horizon for $c_s^2=10^{-7}$ corresponds to much smaller scales and we have an enhancement of perturbations on scales larger than the sound-horizon. For this reason, the non-linear dark energy clustering is much clearer for the simulation with smaller speed of sound. But the enhancement is also present for $c_s^2 = 10^{-4}$. The peak of the dark energy power spectrum is also affected by non-linearities: for example, in the case $c_s^2=10^{-7}$ the peak of the dark energy density power spectrum at redshift $z=0$ is shifted to smaller scales.

\begin{figure}%
    \centering
       \includegraphics[width=1.0\textwidth]{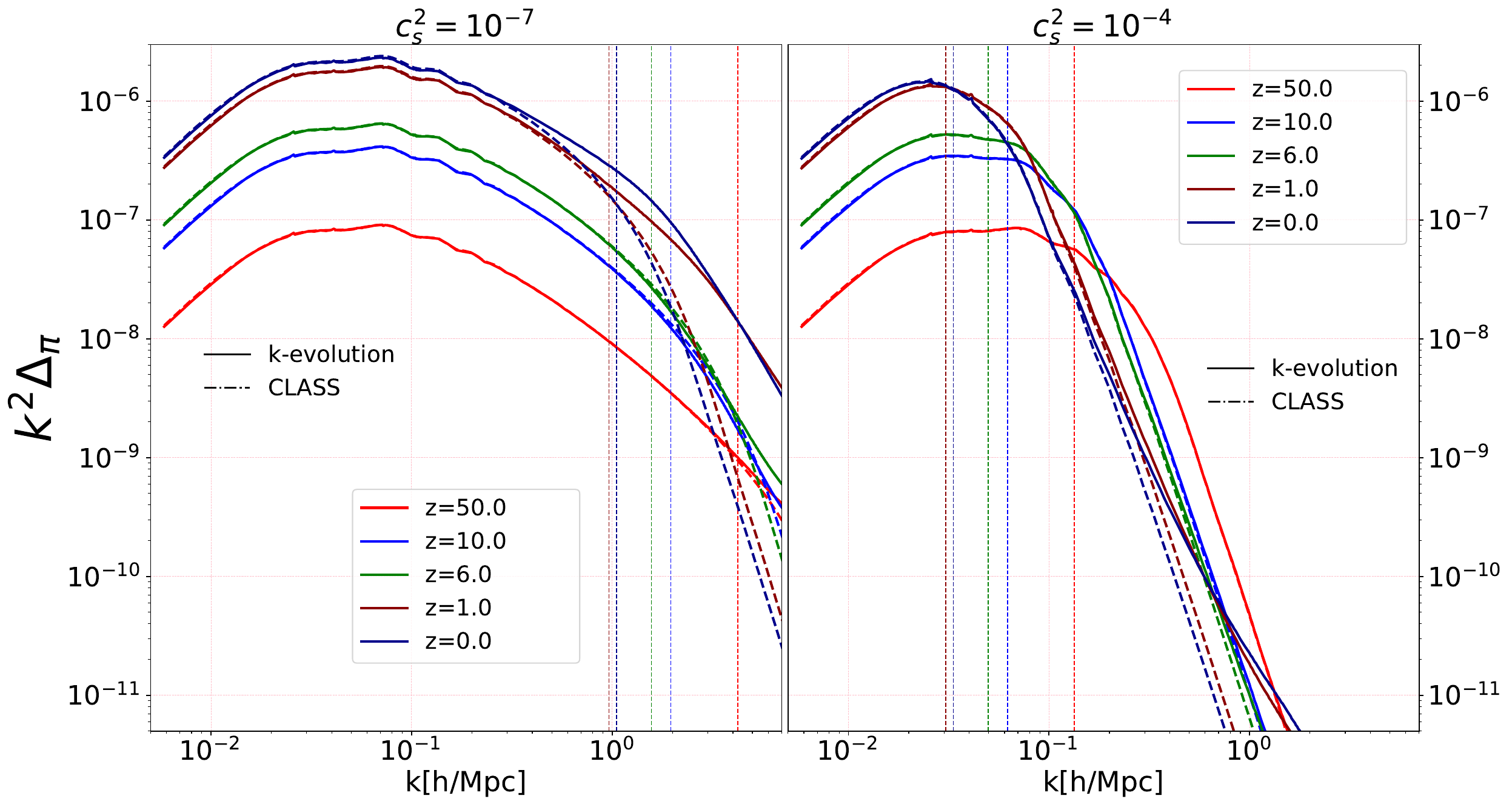}
\caption{Comparison of $\pi$ power spectra at different redshifts from \kev (solid lines) and \class (dashed lines), for $c_s^2=10^{-4}$ (right-panel)  and $c_s^2=10^{-7}$ (left-panel). {$\Delta_{\pi}$ has units of ${\rm Mpc^2}/ h^2$; multiplying it by $k^2$ makes it dimensionless.} The vertical dashed lines show the value of the dark energy sound-horizon at each redshift. The dark energy velocity divergence power spectrum,
 $\Delta_{\theta_{\rm DE}}$, is  simply given by $k^4 \Delta_{\pi}$. }
  \label{fig:pi_class_kev}
\end{figure}
In Fig.~\ref{fig:pi_class_kev}, the {$\pi$} power spectra at different redshifts from \kev and \class for the two speeds of sound are compared. $\Delta_{\pi}$ has units of $[\rm Mpc^2/ \rm h^2]$, multiplying by $k^2$ makes it dimensionless. It is important to note that one can obtain the $\theta_{\rm DE}$ spectrum from the $\pi$ spectrum by using $\pi(k,z) =\theta (k,z)/{k^2}$ according to Eq.\ \eqref{theta_pi}. For both speeds of sound, the dark energy scalar field {fluctuations  $\pi$} or equivalently the dark energy velocity divergence $\theta_{\rm DE}$ becomes non-linear due to the matter non-linearities and decays inside the sound-horizon scale.

 \begin{figure}%
    \centering
       \includegraphics[width=1.0\textwidth]{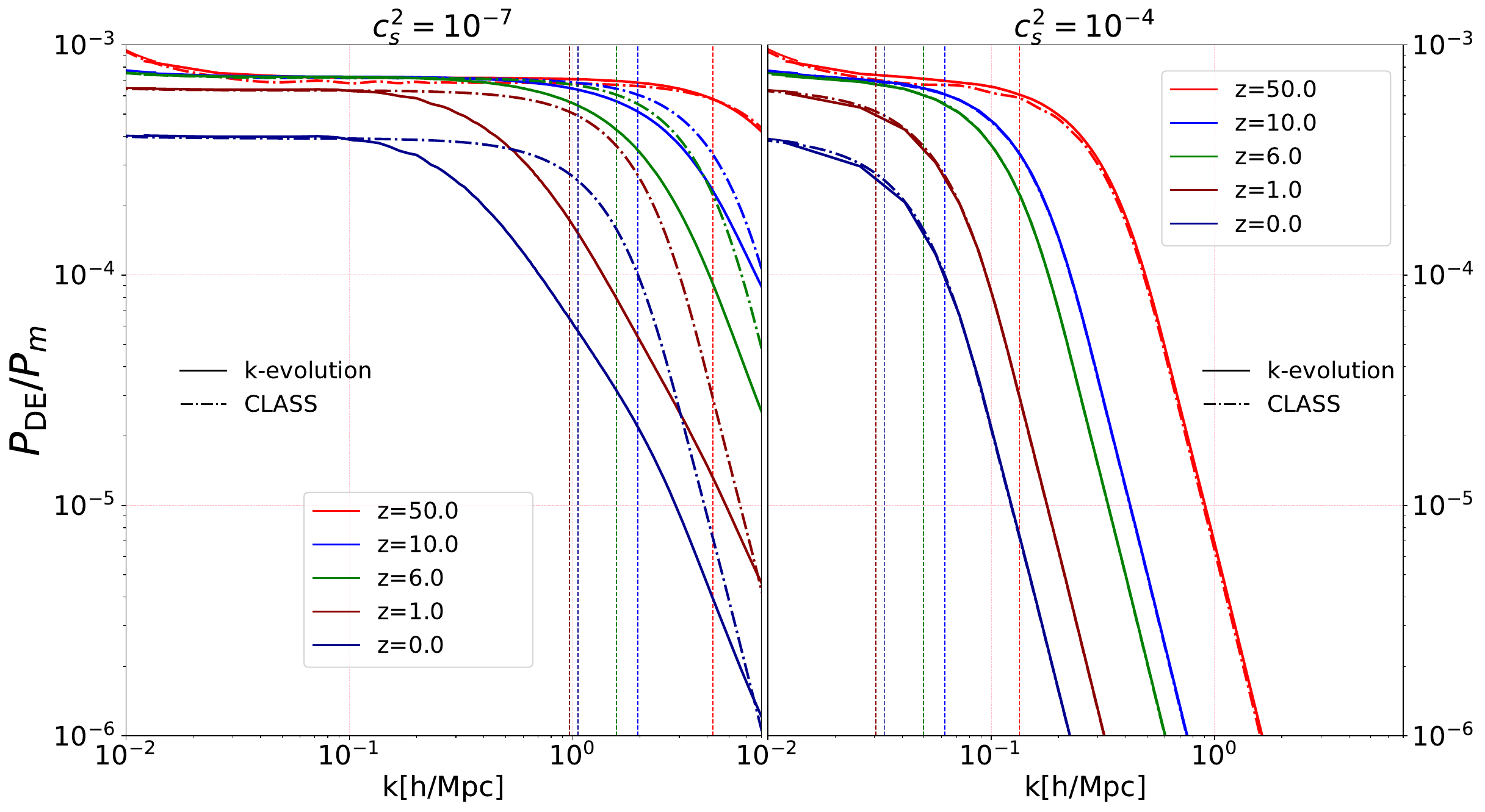}
       \caption{{Comparison of the ratio between dark energy and matter power spectra from \class  and \kev for two speeds of sound, $c_s^2=10^{-4}$ (right panel) and $c_s^2=10^{-7}$ (left panel), at different redshifts. For $c_s^2=10^{-4}$, the sound-horizon is at about the scale of matter non-linearity and the  result from \kev agrees well with the one from \class. For $c_s^2=10^{-7}$, the sound-horizon is smaller than the matter non-linearity scale and we observe significant differences between the \kev and \class results,  due to the matter and dark energy clustering, which are absent in the linear theory.} The upturn visible in the ratio on large scales for $z=50$ is a gauge effect on horizon scales. }
  \label{fig:deltakess_deltam}
 \end{figure}
  {Figure \ref{fig:deltakess_deltam} shows the ratio between the dark energy and  matter power spectra at different redshifts, for both speeds of sound $c_s^2=10^{-4}$ and $c_s^2=10^{-7}.$
{In the case $c_s^2=10^{-4}$, \kev and \class agree well, which shows} that the non-linearity in dark energy and matter {are roughly proportional}. On the other hand, in the case $c_s^2=10^{-7}$ we see a large difference between \class and \kev because the  dark energy sound-horizon lies inside the scale of matter non-linearity. Here,
the ratio $P_{\rm{DE}}/P_{\rm m}$} is more suppressed in \kev than in \class, as the matter non-linearity is more effective than dark energy non-linearity, while inside the sound-horizon ({on the very right of the left-hand panel,} for redshifts, $z=0$ and $z=1$) we see the \kev result becoming larger than the one of \class, due to the effective clustering of  dark energy  at those scales. 
A more detailed study of the power ratio can be found in a companion paper \cite{Hassani:2019}.

\subsection{$k$-evolution versus \textit{gevolution} 1.2 using its \class interface\label{sec:jev}}
 
{Relativistic components that only couple gravitationally to dark matter cluster weakly. It is then a good approximation to describe them using their linear solution.}
 For instance, in $N$-body simulations one can simply include a realisation of the linear density field of such components in the computation of the gravitational potentials. To this end, one first computes the respective linear transfer functions using an Einstein-Boltzmann solver, and then lays down perturbations matching to the random amplitudes and phases that where used as initial data for the simulation.
The correct coupled evolution of dark matter and the additional components is recovered at linear order by construction, and whenever the non-linear growth in the dark matter is completely dominated by its self-gravity one can obtain very accurate results even deep in the non-linear regime. This method has been successfully employed for treating the effect of neutrinos \cite{Brandbyge:2008js,Adamek:2017uiq} or radiation \cite{Brandbyge:2016raj,Adamek:2017grt} on dark matter clustering, {and has been extended to dark {energy} fluids in the current version 1.2 of \textit{gevolution}.}
A conceptually similar implementation has  been recently presented in \cite{Dakin:2019vnj}.

As opposed to $k$-evolution, this method does not allow to track the response of  dark energy  to the gravitational potentials of non-linear matter structures, as illustrated in Fig.\ \ref{fig:codes}. This effect is expected to be relevant in particular for low effective {speed of sound} {of} the fluid, i.e.,\ when the clustering of the dark energy is not strongly suppressed. In this section we study the non-linear matter power spectrum obtained with both methods, which allows us to quantify the accuracy of the simplified linear treatment as implemented in \textit{gevolution}. We demonstrate that the matter power spectra agree extremely well on all scales and at all times, but find some noticeable corrections to the gravitational potential at baryon acoustic oscillations (BAO) scales once dark energy dominates.

   \begin{figure}%
      \centering
       \includegraphics[width=0.95\textwidth]{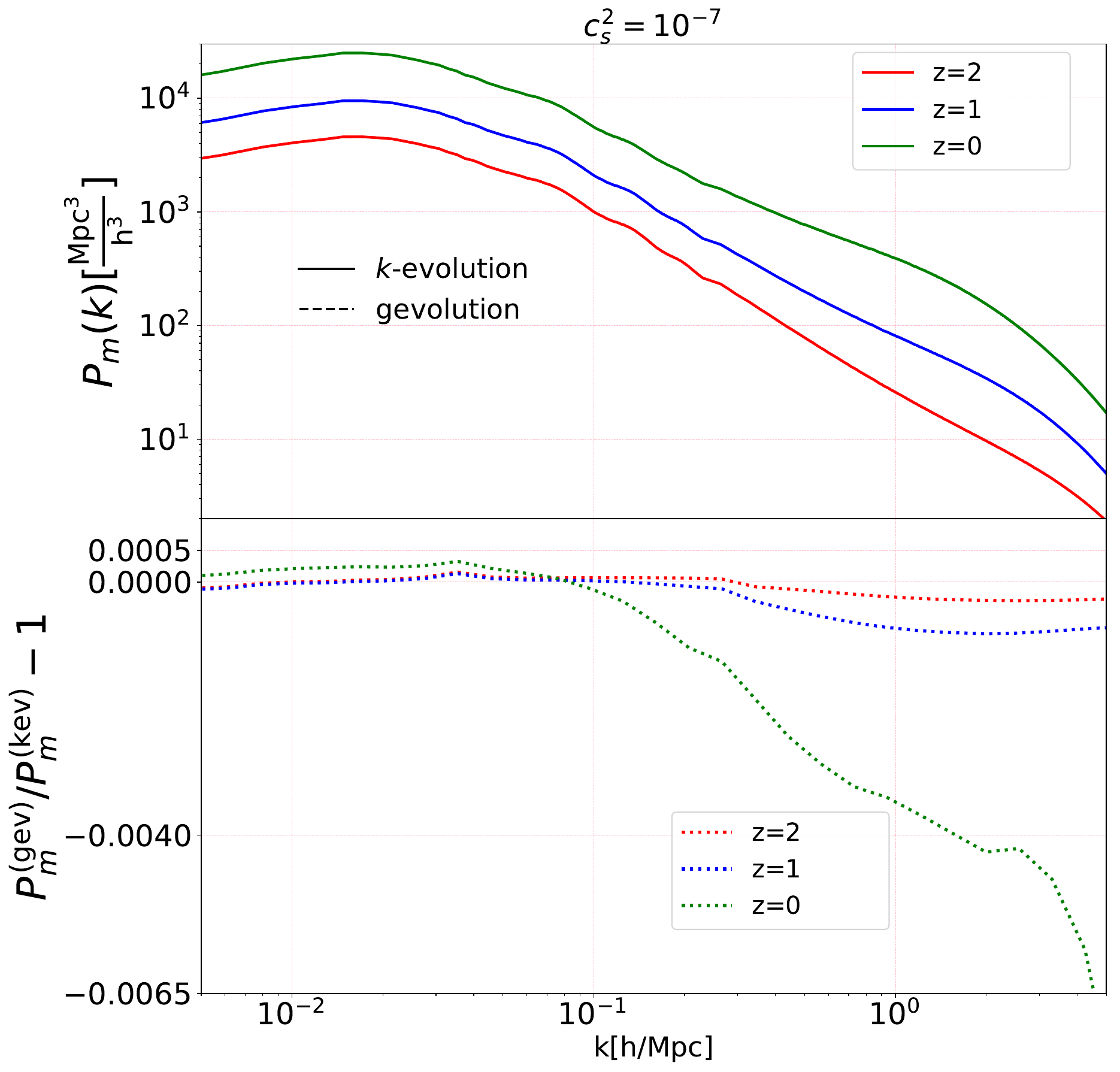}
     \caption{{At the top, comparison of the matter power spectra from \kev and \gev at different redshifts, for $c_s^2=10^{-7}$. At the bottom, relative difference between the two power spectra at each redshift is shown. The figure shows that the effect of non-linearity of dark energy on the matter power spectrum is negligible.}}
  \label{fig:deltam_kev_gev}
 \end{figure}
The matter power spectrum from \kev and \gev at different redshifts are compared in Fig.\ \ref{fig:deltam_kev_gev} for $c_s^2=10^{-7}$, {where} we find a sub-percent agreement on all scales and at all redshifts between the two matter power spectra. For the higher value $c_s^2 = 10^{-4}$ {(not plotted in this figure)} the agreement is even better.
We  conclude that there is no significant impact of {non-linear} dark energy {fluctuations} on the matter spectrum, once non-linear {matter} clustering is correctly taken into account.

  \begin{figure}%
    \centering
           \includegraphics[width=0.95\textwidth]{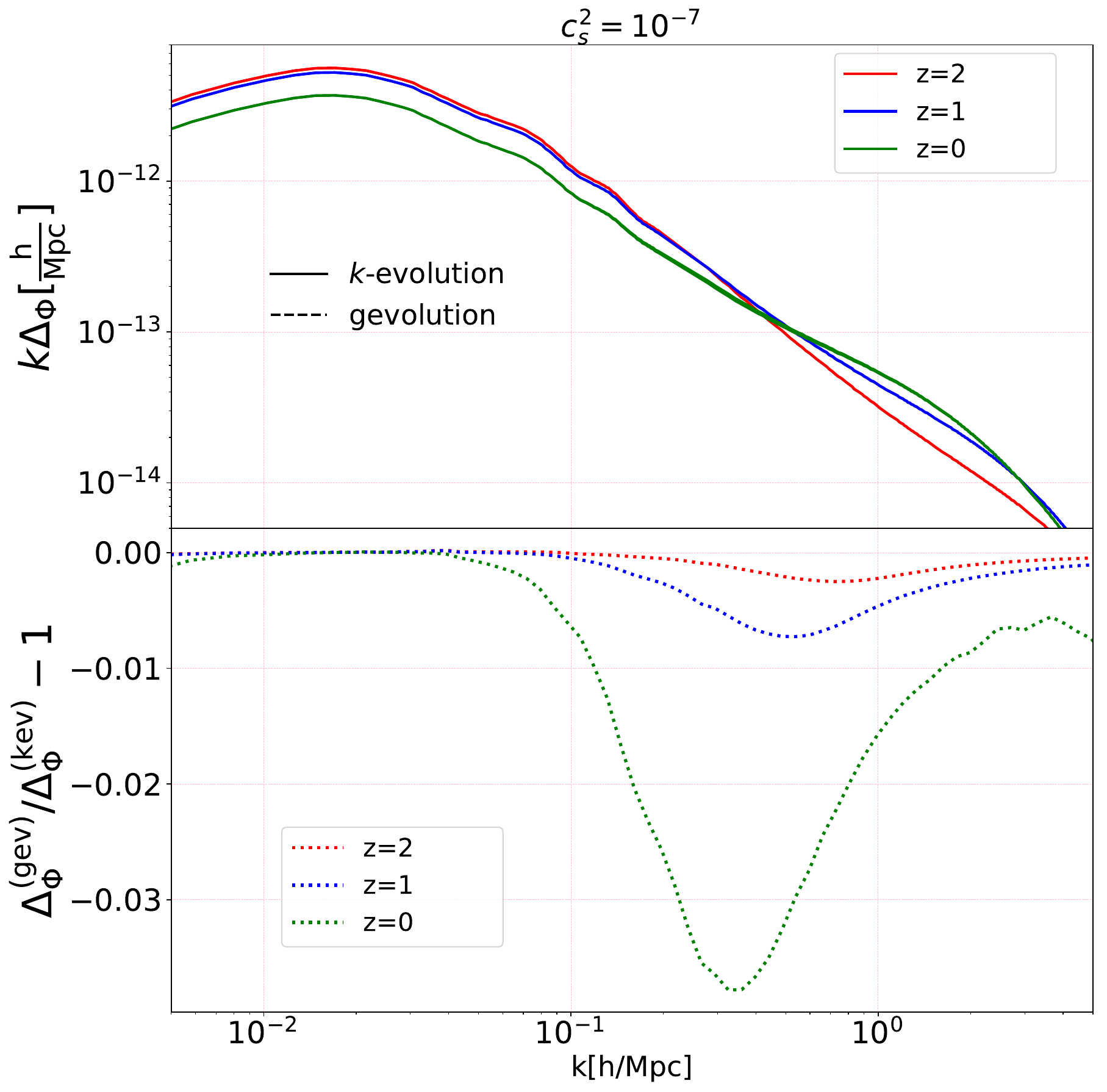}

     \caption{In the top panel, the potential power spectra from \kev and \gev at three different redshifts are shown and in the bottom panel the relative difference between the power spectra at the same redshifts are plotted. The dark energy clustering changes the potential power spectrum by up to $\sim 4 \%$ at mildly linear scales at $z=0$.}
  \label{fig:phi_kev_gev}
 \end{figure}
In Fig.\ \ref{fig:phi_kev_gev} the gravitational potential power spectra from \kev and \gev at different redshifts for the speed of sound $c_s^2=10^{-7}$ are compared. Interestingly, the dark energy clustering affects the gravitational potential power spectrum in the mildly non-linear regime, {with up to $\sim 4\%$ differences appearing at $z=0$. This} effect could potentially change the lensing signal if the universe {contains} $k$-essence as a dark energy fluid.

 \begin{figure}%
    \centering
\includegraphics[width=\textwidth]{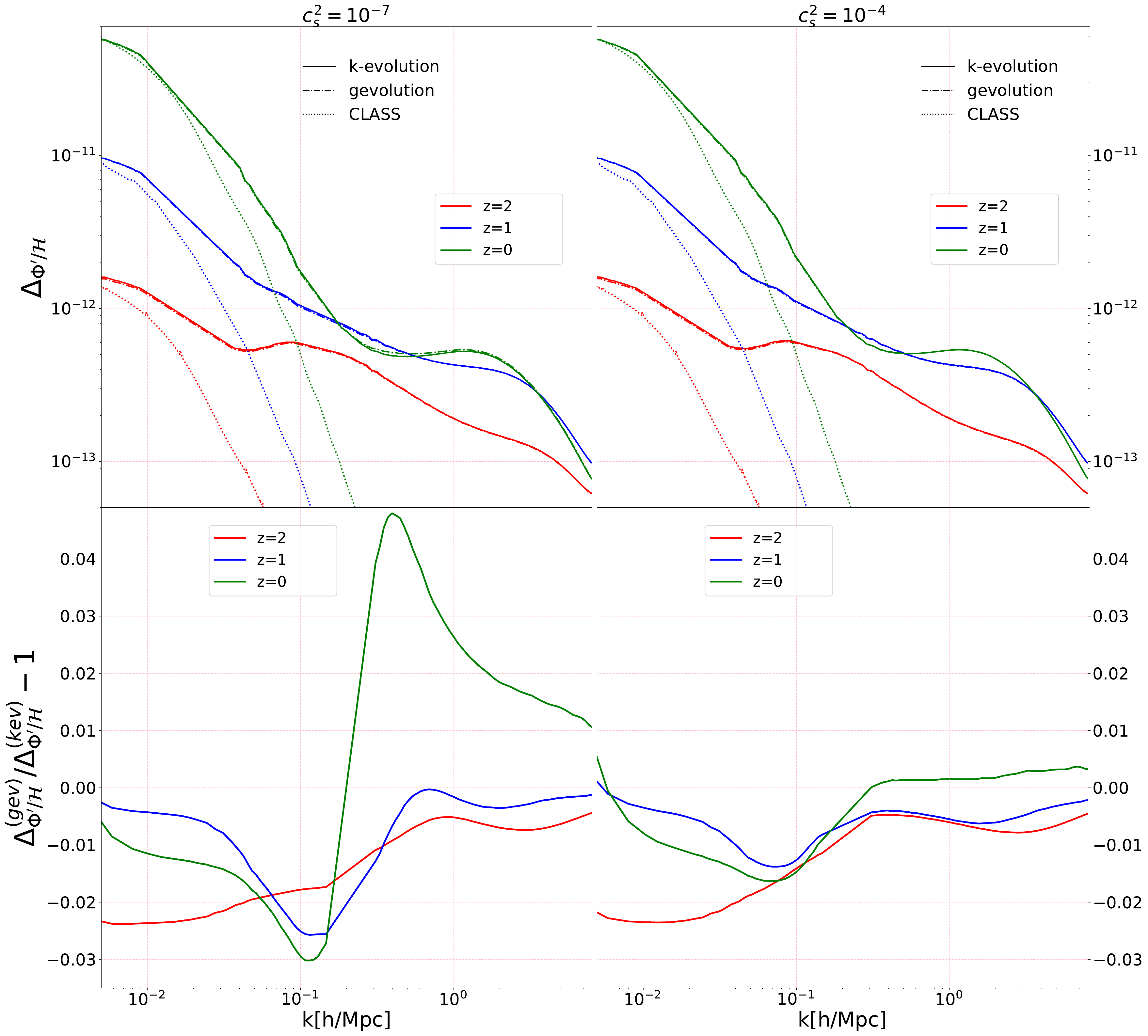}
     \caption{Comparison of the power spectra of  $\Phi'/\HH$  for \kev, \gev\ and \class at different redshifts, for $c_s^2=10^{-4}$  (right panel) and $c_s^2=10^{-7}$ (left panel). The onset of non-linear effects in $\Phi'$ is on much larger scales than in $\Phi$ or $\delta_m$. The results from \kev\ and \gev\ are more similar, with differences reaching to $\sim$5\% between the two codes.
 }
  \label{fig:phi_prime_kev_gev}
 \end{figure}
The power spectra of the dimensionless quantity $\Phi'/\HH$  from \kev and \gev, compared to the linear prediction from \class are shown in the top panel of Fig.\ \ref{fig:phi_prime_kev_gev}.  
This is an interesting quantity as it is a direct source of  dark energy already at linear order, as can be seen in Eq.~\eqref{zeta_eq2}. According to Fig.\ \ref{fig:phi_prime_kev_gev}, the scale of non-linearity in the $\Phi'$ power spectrum starts at  $\sim 0.005 \; h/ \rm{Mpc}$, much {earlier} than the scale of non-linearity in the matter and potential power spectra, which at $z=0$ is $\sim 0.1 \, h/\rm{Mpc}$. This effect is not specific to $k$-essence models, it has also been observed for $\Lambda$CDM in \cite{Cai:2008sm}, where it was studied as non-linear integrated Sachs–Wolfe (ISW) effect. Interestingly, this effect adds a new scale into the dynamics of {$\pi$}, in addition to the sound-horizon scale and the scale of matter non-linearity. Comparing \kev and \gev, there is a $\sim 3\%$ effect due to the clustering of dark energy 
at large and quasi-linear scales for both {values of $c_s^2$}. Moreover, there is $\sim 5 \%$ {bump} due to the dark energy clustering at quasi-linear scales for the case $c_s^2=10^{-7}$, which peaks around $k=0.4 \, h/\rm{Mpc}$.} 
These changes in $\Phi'$, and especially the large difference relative to the linear predictions, even at low $k$, could potentially affect the ISW effect. 
 \begin{figure}%
    \centering
\includegraphics[width=\textwidth]{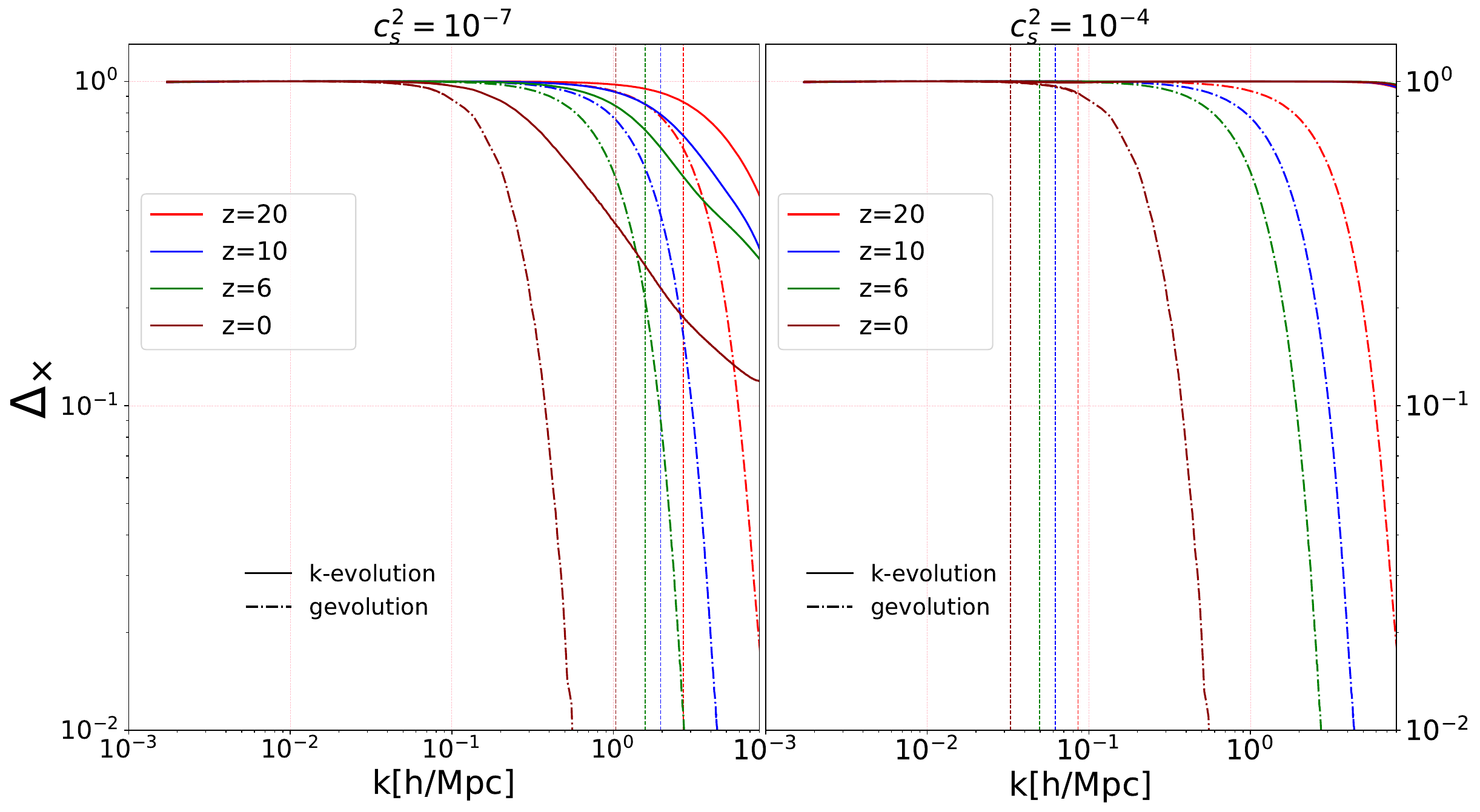}
     \caption{Normalized cross correlation power spectra at different redshifts for speed of sound $c_s^2=10^{-4}$ on the right and $c_s^2=10^{-7}$ on the left. In \gev, for both speeds of sound the only important scale is the scale of matter non-linearity as in \gev the dark energy  does not follow the non-linear matter structures, and we see that after this scale the cross correlation power decays. In \kev for $c_s^2=10^{-4}$ almost at all scales the dark energy and matter densities are fully correlated, as inside the sound-horizon  dark energy does not cluster strongly and closely follows the matter density. In the case with lower speed of sound, where dark energy clusters and has self-dynamics the cross correlation power starts to decay on small scales.}
  \label{fig:cross_kev_gev_normal}
  \end{figure}
  
{We define the normalized cross power spectrum between matter and dark energy as
\begin{equation}
\Delta_\times = \frac{\Delta_{\rm{DE} \times \rm{m}}}{{\sqrt{\Delta_{\rm DE} \Delta_{\rm m}}}} \, ,
\end{equation}
where $\Delta_{\rm DE}$ and $\Delta_{\rm m}$ are respectively the dark energy and the matter power spectrum while $\Delta_{\rm{DE} \times \rm{m}}$ is their cross  spectrum. This quantifies the correlation between the clustering of matter and dark energy. 
Figure \ref{fig:cross_kev_gev_normal} compares this quantity computed with $k$-evolution and \gev.
In particular, it shows the cross-spectra at different redshifts, for $c_s^2=10^{-4}$ (right panel) and $c_s^2=10^{-7}$ (left panel).

According to the Cauchy-Schwarz inequality, this quantity must be in the range $[-1,1]$. A value of 1 indicates that the two fields are fully correlated, 0 means that they are not correlated, and -1 that they are fully anticorrelated. In linear perturbation theory and for adiabatic initial conditions (assumed here), all quantities are related via a deterministic transfer function to the same initial curvature perturbation, so that all the fields are fully correlated, $
\Delta_\times = 1$.}

We see that on large scales, where the evolution is effectively linear, the dark energy and matter fluctuations are indeed fully correlated in all cases. In \gev, the matter evolves non-linearly under its own gravity, while dark energy is computed at the linear level. For this reason the two fields start to lose their correlation when the matter perturbations become non-linear.

{In \kev\ on the other hand, where  dark energy  is able to follow the dominant non-linear matter perturbations, the correlations are essentially maintained for large {speeds of sound} such as $c_s^2 = 10^{-4}$. In this case,  dark energy  crosses its sound-horizon before the scale of matter non-linearity and it is not able to develop an independent dynamics; its clustering simply follows that of the dark matter.}

The situation is different for low speeds of sound, such as $c_s^2=10^{-7}$, where dark energy becomes non-linear outside the sound-horizon. In that case the correlations start to decay, but more slowly than in \gev, as the matter clustering is still dominant and ``drags'' the dark energy perturbations at least partially with it. This behaviour is clearly visible also in the field snapshots that we study in Sec.~\ref{sec:snapshots}.

  \begin{figure}%
    \centering
\includegraphics[width=\textwidth]{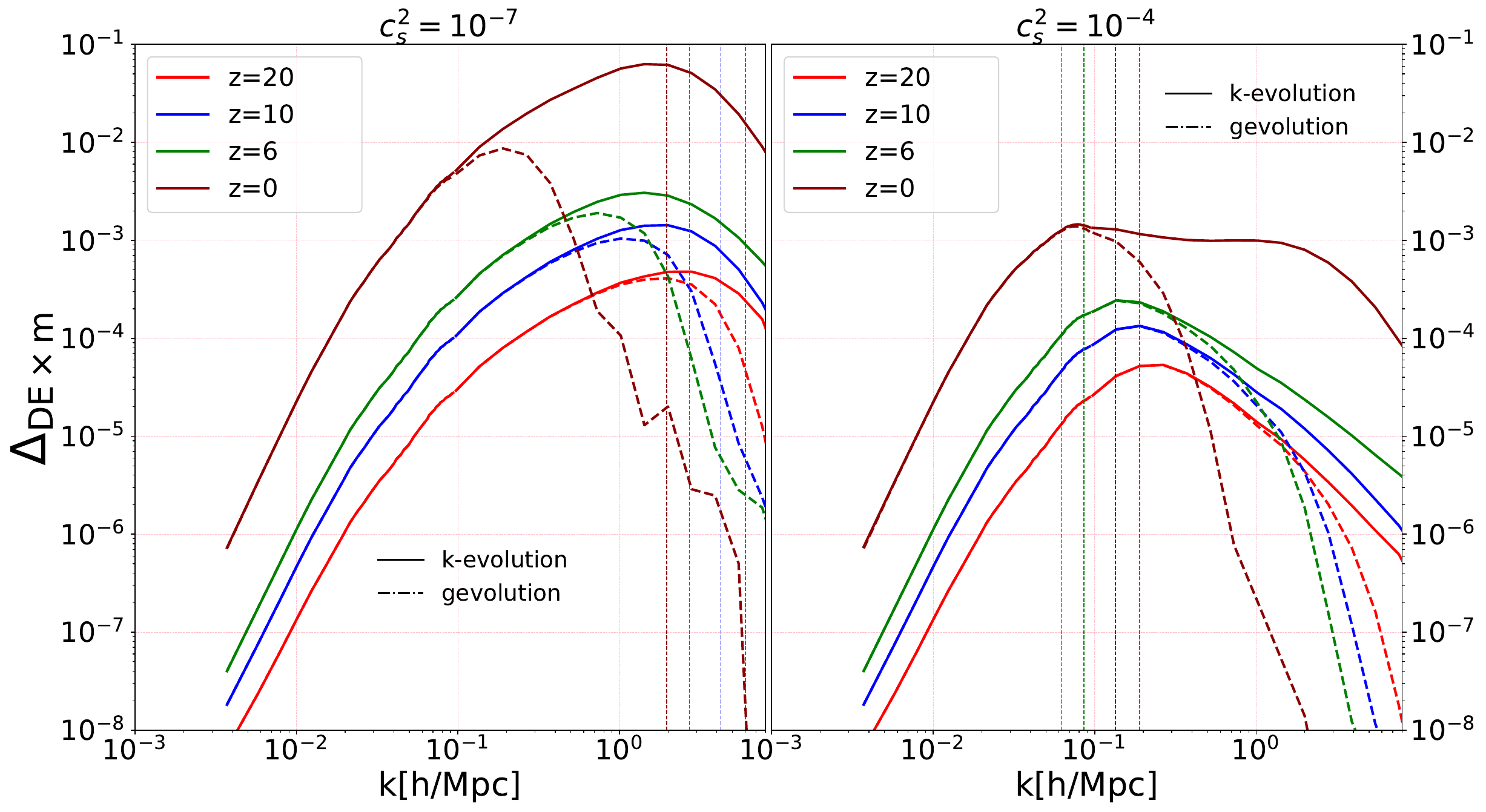}
     \caption{{Cross-correlation power spectra between matter and dark energy densities at different redshifts, for $c_s^2=10^{-4}$ (right panel) and $c_s^2=10^{-7}$ (left panel), computed by $k$-evolution and \gev.  
     In \gev, the dynamics of dark energy and matter decouple beyond the scale of matter non-linearity, so the dashed lines and solid lines start deviating roughly at the scale of non-linearity. In \kev, for  $c_s^2=10^{-4}$ the dark energy density does not cluster at scales where matter clusters, and the dark energy follows matter. This is why the cross-correlation power between the two densities is large. For $c_s^2=10^{-7}$, the dark energy density clusters and the turn-around in the cross-correlation power spectrum takes place at the sound-horizon scale.} }
  \label{fig:cross_kev_gev}
  \end{figure}
  For completeness we also show the raw cross-spectrum between matter and dark energy densities at different redshifts for both speeds of sound in Fig.\ \ref{fig:cross_kev_gev}. The main feature is the enhanced cross-power on small scales in \kev due to the non-linear clustering of  dark energy, mostly following the non-linear dark matter clustering.
  \begin{figure}%
    \centering
\includegraphics[width=\textwidth]{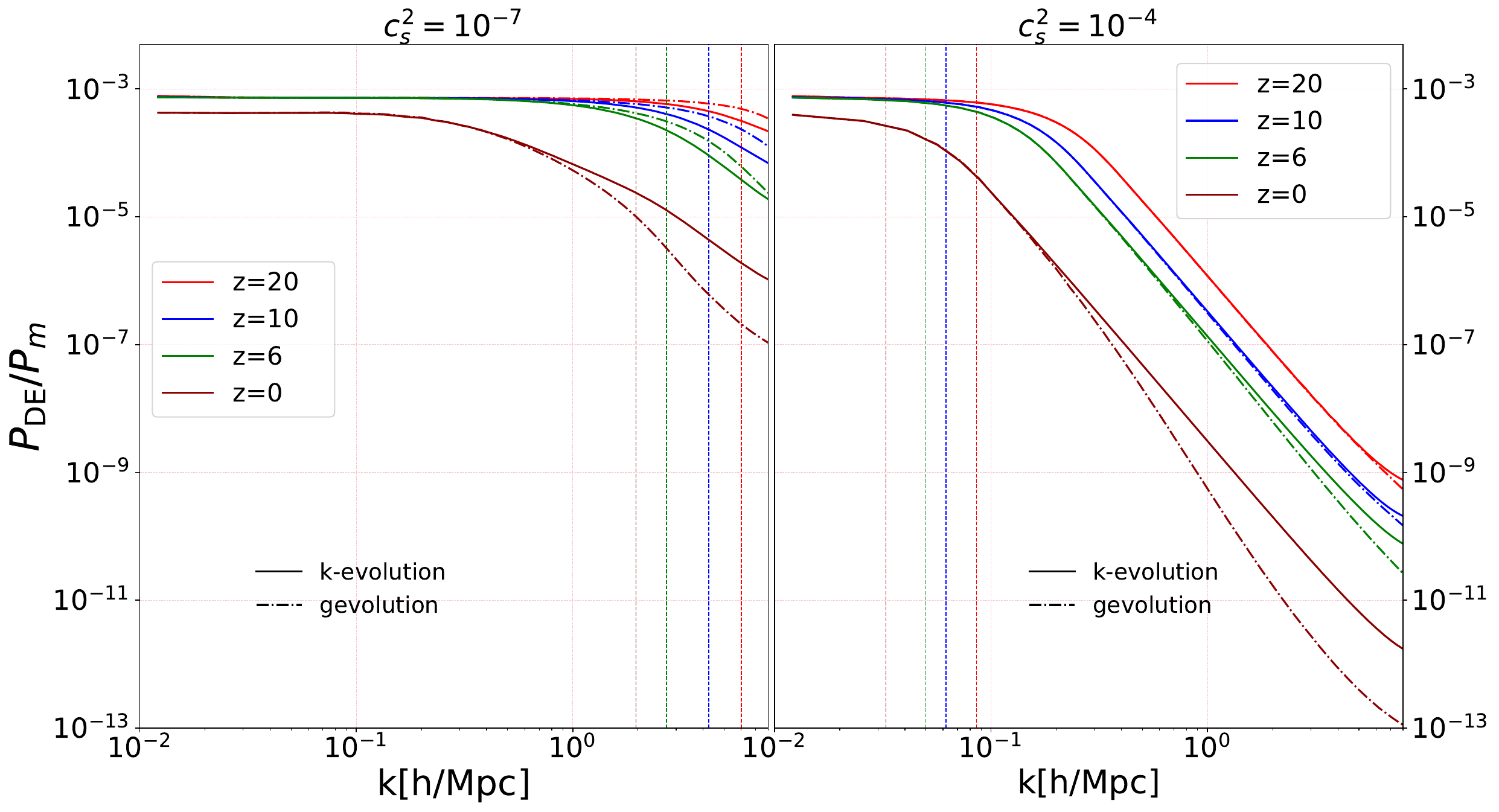}
     \caption{Ratio of dark energy and matter densities power spectra in \kev (solid lines) and \gev (dashed lines), at different redshifts for $c_s^2=10^{-7}$ (left panel) and $c_s^2=10^{-4}$ (right panel). The scale at which  
     \gev and \kev start disagreeing is giving by a combination of sound-horizon and scale of matter non-linearity. In the low speed of sound case, $c_s^2=10^{-7}$, this happens at smaller scales. Also this ratio increases with the redshift, as the matter clustering is much stronger than the dark energy clustering at lower redshifts.}
  \label{fig:ratio_densities_gev_kev}
 \end{figure}
  In Fig.\ \ref{fig:ratio_densities_gev_kev} the ratio of dark energy and matter power spectra at different redshifts are compared. On scales above the sound-horizon we expect during matter domination a ratio of \cite{Creminelli:2008wc,Sapone:2009mb}
\begin{equation}
\frac{P_{\rm DE}}{P_m} \simeq \left( \frac{1+w}{1-3 w} \right)^2 = \frac{1}{1369}\;,
\end{equation}
for $w=-0.9$, which is verified by the simulations. Dark energy perturbations inside the sound-horizon stop growing, so that the ratio relative to the dark matter perturbations decreases. At lower redshifts, the decrease in $k$-evolution tends to be slower than in \gev, since in the latter only the dark matter perturbations become non-linear on small scales, while dark energy is always linear. In \kev both dark matter and dark energy perturbations become non-linear on small scales.
\subsection{Newtonian simulations with ``back-scaled'' initial conditions}
{Dark energy is a key target for large future surveys like the ESA Euclid satellite \cite{Laureijs:2011gra}, and to exploit such data fully it is necessary to have reliable results also on small scales where the matter perturbations are non-linear.} To find these results, it is common to use Newtonian $N$-body simulations where only the expansion rate is changed, and where no dark energy  is included \cite{Jennings:2009qg}. However, although the dark energy perturbations are small, it is not clear whether this is really a good approximation as we know that the CMB temperature anisotropies on large scales are very sensitive to the perturbations \cite{Weller:2003hw}. 
In order to study this question, we compare this standard approach with our method that includes the linear dark energy perturbations, as implemented in \textit{gevolution} 1.2.

At the linear level, the presence of dark energy perturbations induces a scale dependence in the growth of matter perturbations. This means that simulations where only the background evolution is adjusted do not even reproduce the linear results correctly. In order to deal with this issue, the common practice is to first choose a redshift at which the accuracy of the $N$-body simulation should be maximal (e.g.,\ redshift zero) and to compute a linear matter power spectrum for that redshift, including the effects of dark energy perturbations. In a second step, the matter power spectrum is then ``scaled back'' to the initial redshift of the simulation with the scale-independent growth function obtained by neglecting the dark energy perturbations, based only on the modified expansion rate. While this procedure provides initial data that do not correspond to the true matter configuration at the initial time, the error is deliberately introduced in order to precisely cancel the error in the linear evolution once the simulation reaches the final redshift.

In our comparison we consider a case with a low speed of sound, $c_s^2 = 10^{-7}$. In principle such a choice may pose a challenge to the standard approach, as a scale dependence is introduced close to the non-linear scale where the procedure outlined above becomes less reliable. We run one simulation with \textit{gevolution} using the  {\class interface} to provide our baseline, and then compare our results with two simulations where the {\class interface} is not used and only the background evolution tracks the dark energy equation of state. We provide ``back-scaled'' initial conditions for the latter two simulations, in one case based on the correct linear matter power at redshift $z=0$, while in the other case we match to the linear matter power when additionally $c_s^2 = 1$ is assumed for the dark energy, as is done in most numerical studies. All simulations are run in the Newtonian mode, which means that for the baseline simulation the fluid perturbations are taken in the $N$-body gauge \cite{Fidler:2016tir}.

\begin{figure}[t]
\includegraphics[width=0.95\textwidth]{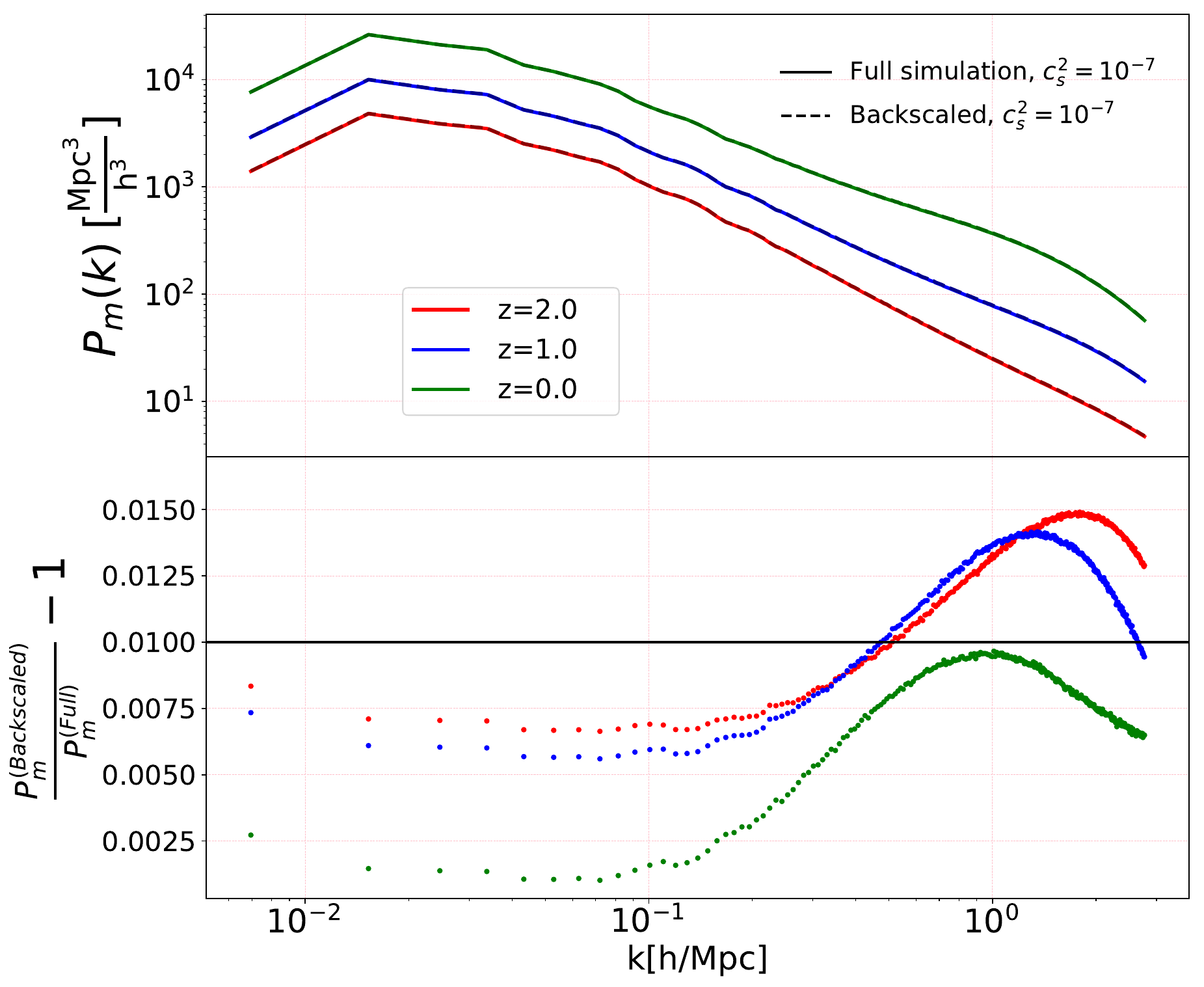} 
\caption{The matter power spectra and the relative difference from two Newtonian simulations, one with the back-scaled initial conditions for $c_s^2 =10^{-7}$ and one with the correct initial conditions from the linear Boltzmann code \class and correct evolution for $c_s^2 =10^{-7}$ {including linear dark energy perturbations}. The simulation with back-scaled initial condition works well at $z=0$ especially at linear scales by construction while it reaches 1\% {error} at non-linear scales. At higher redshifts the relative errors are typically larger, but remain below 2\% on all scales.
}
\label{fake_deltam_cs7}
\end{figure}

In Fig.~\ref{fake_deltam_cs7} we compare the 
matter power spectra of a simulation that used the ``back-scaling'' approach with the baseline simulation. As expected, Fig.~\ref{fake_deltam_cs7} shows that the linear scales agree to high accuracy at $z=0$ when the initial conditions are constructed appropriately. However, the effect of scalar field clustering on the matter power spectrum reaches $\sim$1\% at small scales at $z=0$, if the initial conditions are prepared for $c_s^2=10^{-7}$.
{As back-scaled Newtonian simulations are often performed with a quintessence-motivated spectrum where $c_s^2=1$, we also performed a back-scaled simulation with this wrong speed {of sound}. We found that this choice does not have a large impact on the results; the errors on small scales are even reduced relative to correct back-scaled case. This might be  induced because the large-scale and small-scale errors have opposite sign at earlier redshifts.}
  
\begin{figure}[t]
 \includegraphics[width=0.95\textwidth]{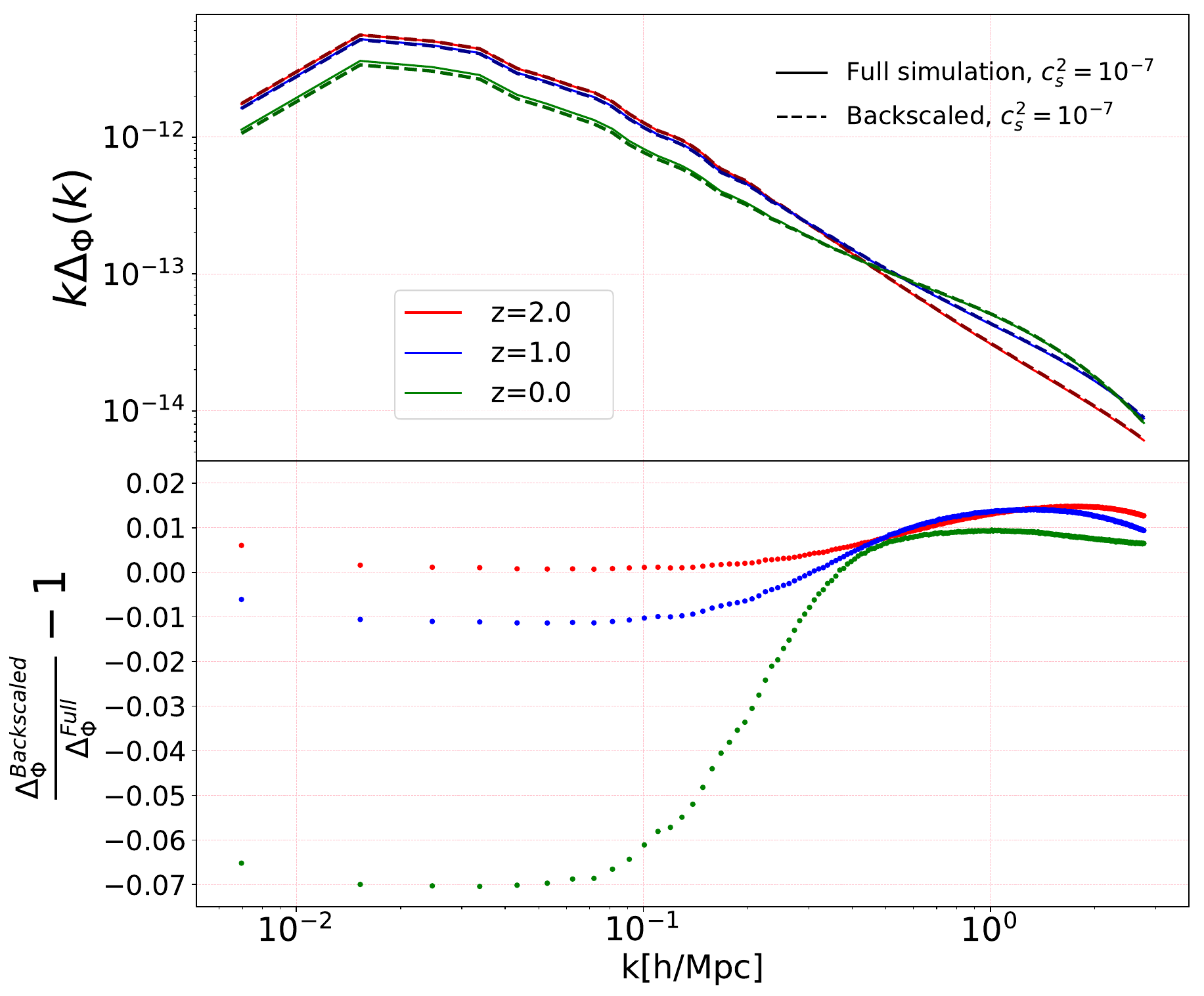} 
 \caption{The gravitational potential power spectra and the relative difference from two Newtonian simulations, one with the back-scaled initial conditions for $c_s^2 =10^{-7}$  and one with the correct initial conditions and correct evolution for $c_s^2 =10^{-7}$ {including linear dark energy perturbations}. The simulation with back-scaled initial condition does not give the {correct} gravitational potential power spectrum at large scales; we find about $7 \%$ relative {error}.}
  \label{fake_phi_cs7}
 \end{figure}

The gravitational potential comes with much larger deviations, as is shown in Fig.~\ref{fake_phi_cs7}. 
The relative difference in the respective power spectra reaches 7\% and could affect the lensing signal. The reason for this large deviation is that back-scaled initial conditions are constructed to produce an accurate matter density but not gravitational potential. The latter is additionally sourced by the perturbations in the dark energy  which, however, are only important at late times. {It would be possible instead to use initial conditions that improve the agreement for the gravitational potential, but then the matter power spectrum would be off. Using the back-scaling approach it is not possible to obtain good results for both the matter density and the gravitational potential simultaneously. In a related paper  \cite{Hassani:2019} we present a possible approach to include a correction for the gravitational potential that addresses this problem.} 

\section{Snapshot analysis\label{sec:snapshots}}

\begin{figure}[hbt!]
  \centering
  \begin{subfigure}[b]{0.95\textwidth}
    \includegraphics[width=\textwidth]{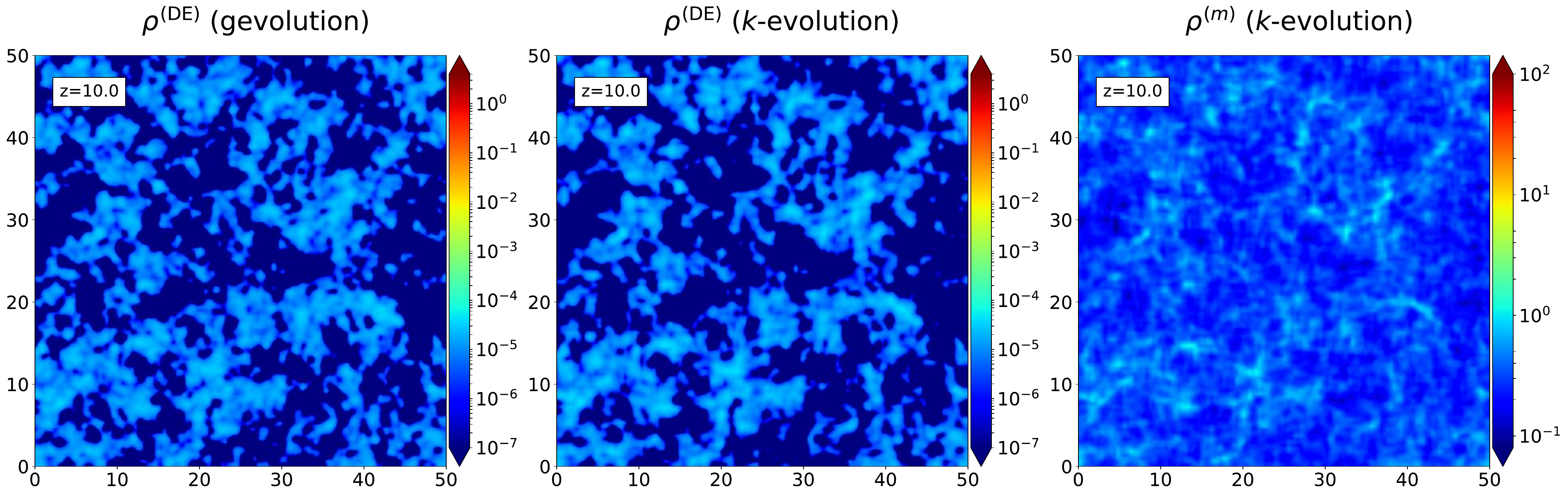}
  \end{subfigure}
  \begin{subfigure}[b]{0.95\textwidth}
    \includegraphics[width=\textwidth]{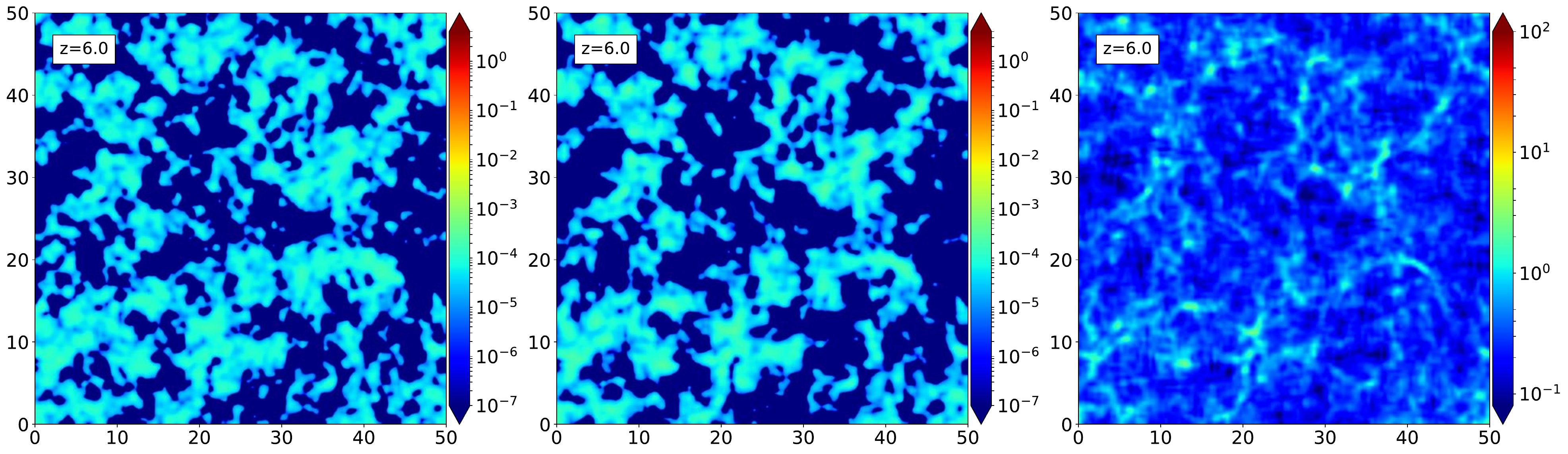}
  \end{subfigure}
  \begin{subfigure}[b]{0.95\linewidth}
    \includegraphics[width=\linewidth]{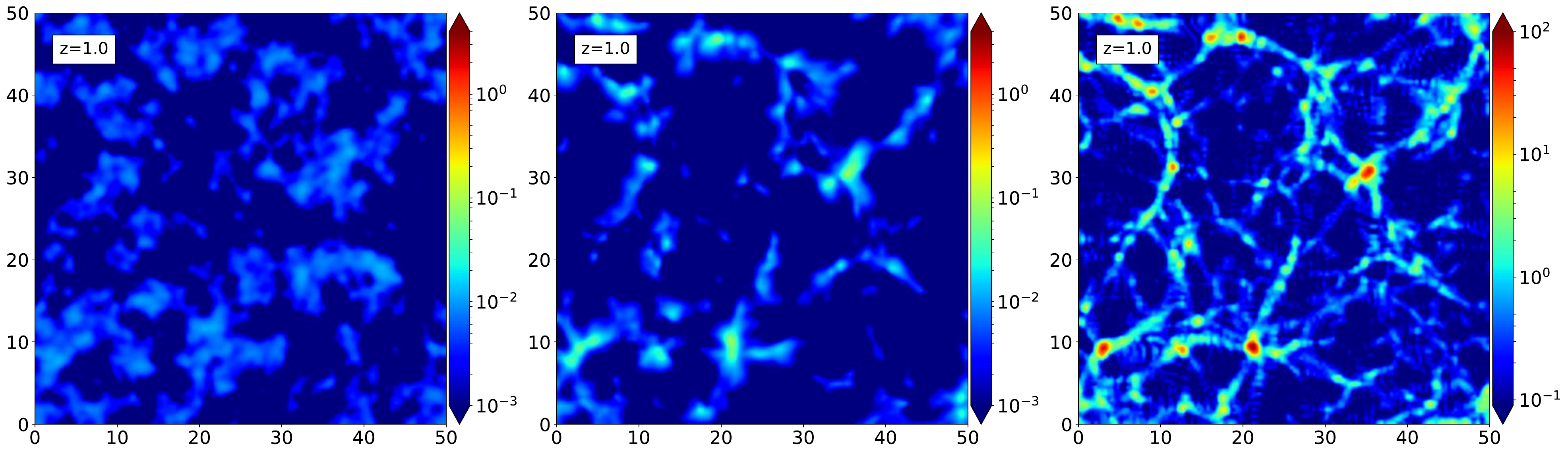}
  \end{subfigure}
  \begin{subfigure}[b]{0.95\linewidth}
    \includegraphics[width=\linewidth]{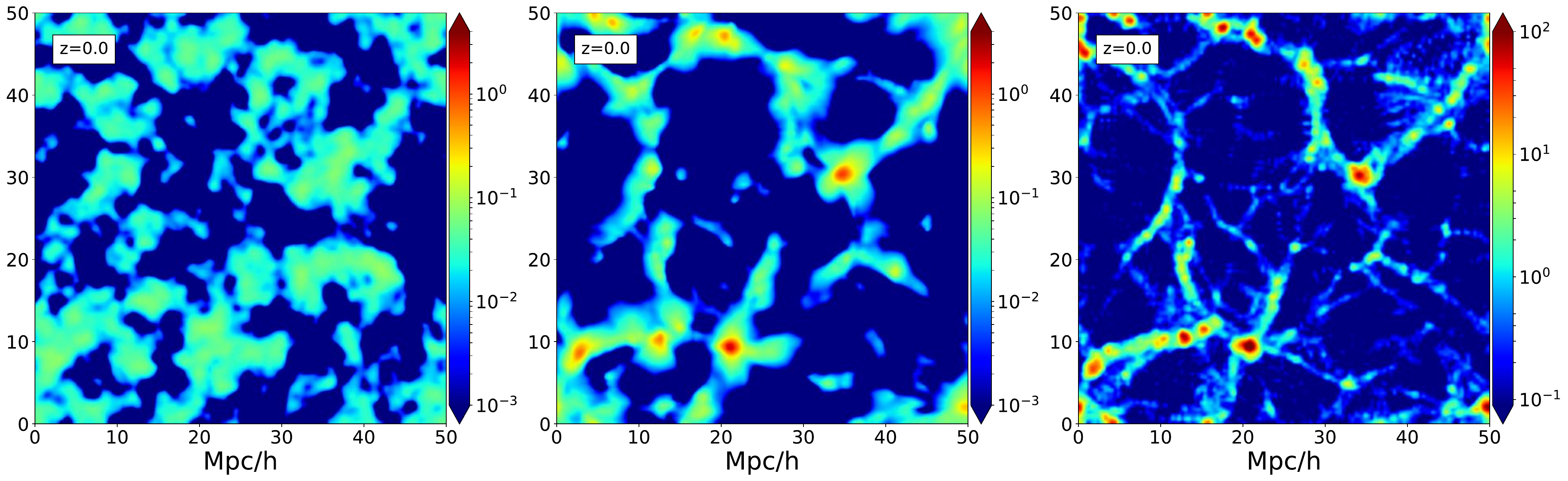}
  \end{subfigure}
  \caption{{Respectively from the left to the right, the dark energy density computed with \gev (using the \class interface) and the dark energy and matter density computed with $k$-evolution, as a function of the redshift (from the bottom to the top), measured in units of the critical density. The dark energy structures form around massive halos. Note that the color scheme for the visualisation of $\rho^{\rm (DE)}$ changes between lower  panels and upper panels.}}
  \label{fig:kess_evolution}
\end{figure}

In this section we look at the matter and dark energy   density from the $k$-evolution and {\textit{gevolution}} simulations (where the latter uses the \class interface), to study how the dark energy field {configuration traces} the matter structures. First, we compare the results obtained from $k$-evolution and {\textit{gevolution}} for a relatively small simulation (128{$^\mathrm{3}$ grid points})
with a small box (50 Mpc$/h$), which corresponds to {a spatial} resolution {of} $~0.39$ Mpc/$h$ and {a} mass resolution {of} $5 \times 10^{9} \, \textup{M}_{\odot}/h$. In Fig.~\ref{fig:kess_evolution}, {we show a 2D slice of   the box, which passes through} 
the most massive halo found by the {ROCKSTAR} halo finder {\cite{Behroozi:2011ju}.}{ Comparing the left and middle panels of Fig.\ \ref{fig:kess_evolution} we see that at high redshift, $z=10$, they are virtually indistinguishable, which is still nearly true at $z=6$. At low redshifts, $z=1$ and $z=0$, the \kev results are clearly more clustered than the linear dark energy realisation of \gev. The dark energy clustering is most pronounced in regions of strong dark matter clustering, i.e.\  {dark energy structures are formed around massive dark matter halos, something that is not the case in the linear realisation.} This agrees with the relatively high correlation between dark matter and dark energy perturbations visible in Fig.\ \ref{fig:cross_kev_gev_normal}.}

 \begin{figure} [hbt!]
 \centering
   \begin{subfigure}[b]{1.0\textwidth}
 \includegraphics[scale=0.225]{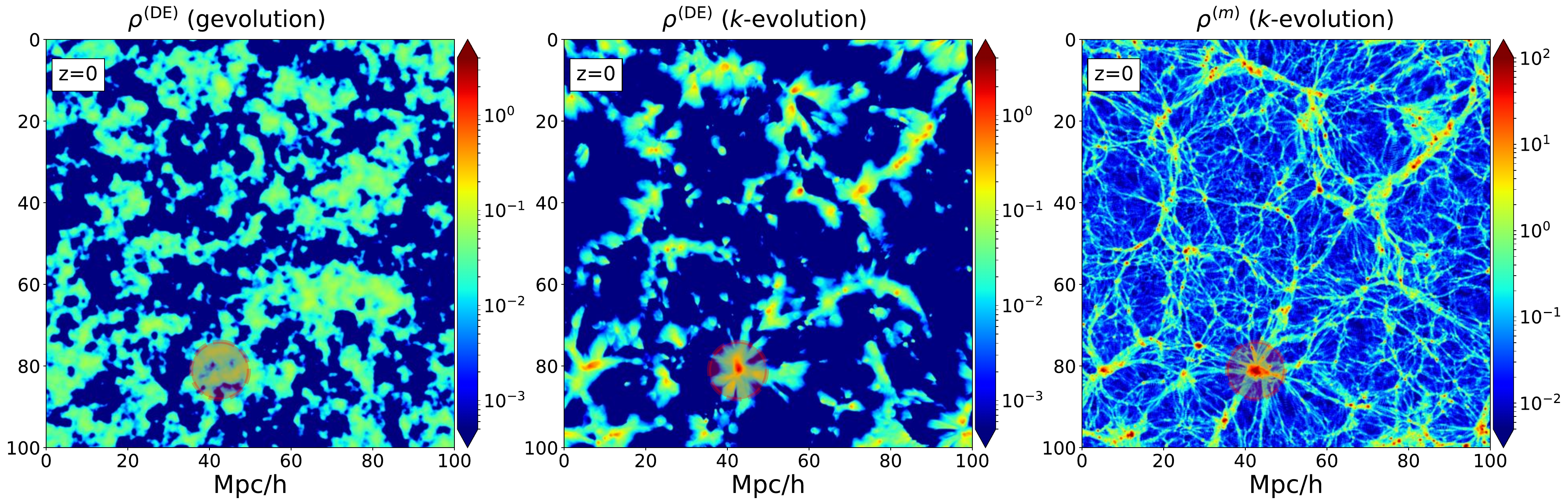} 
 \caption{ Snapshot for dark energy density from \gev (left panel), \kev (middle panel) and matter density from \kev(right panel)  measured in the critical density unit at $z=0$ from a high resolution simulation is shown. The shaded region shows the most massive halo in the simulation which is going to be studied in detail in the next figures. }
 \label{halos_po}
  \end{subfigure}
     \begin{subfigure}[b]{1.0\textwidth}
      \hbox{\hspace{0.1em} 
 \includegraphics[scale=0.225]{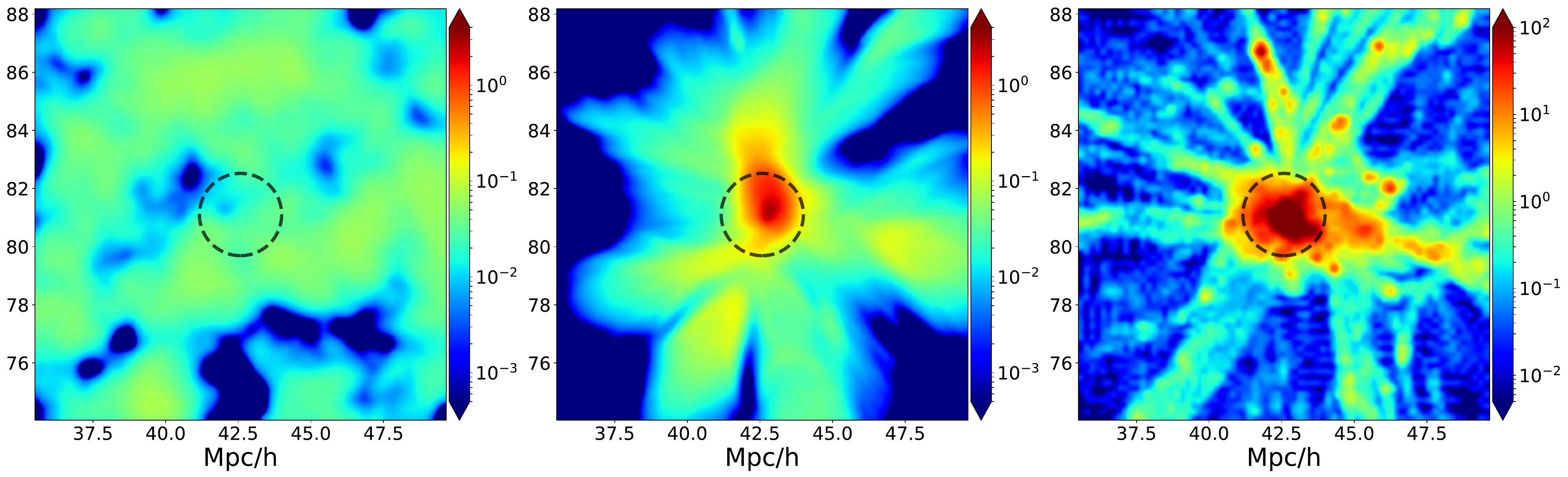} }
 \caption{A close look at the most massive halo (shaded region in the previous snapshot). The color bar range for dark energy density is different with matter density, as dark matter clusters more efficiently than dark energy. }
 \label{halo_close_look}
  \end{subfigure}
 
  \centering
    \begin{subfigure}[b]{1.0\textwidth}
\centering
\hbox{\hspace{-0.04em} 
 \includegraphics[scale=0.24]{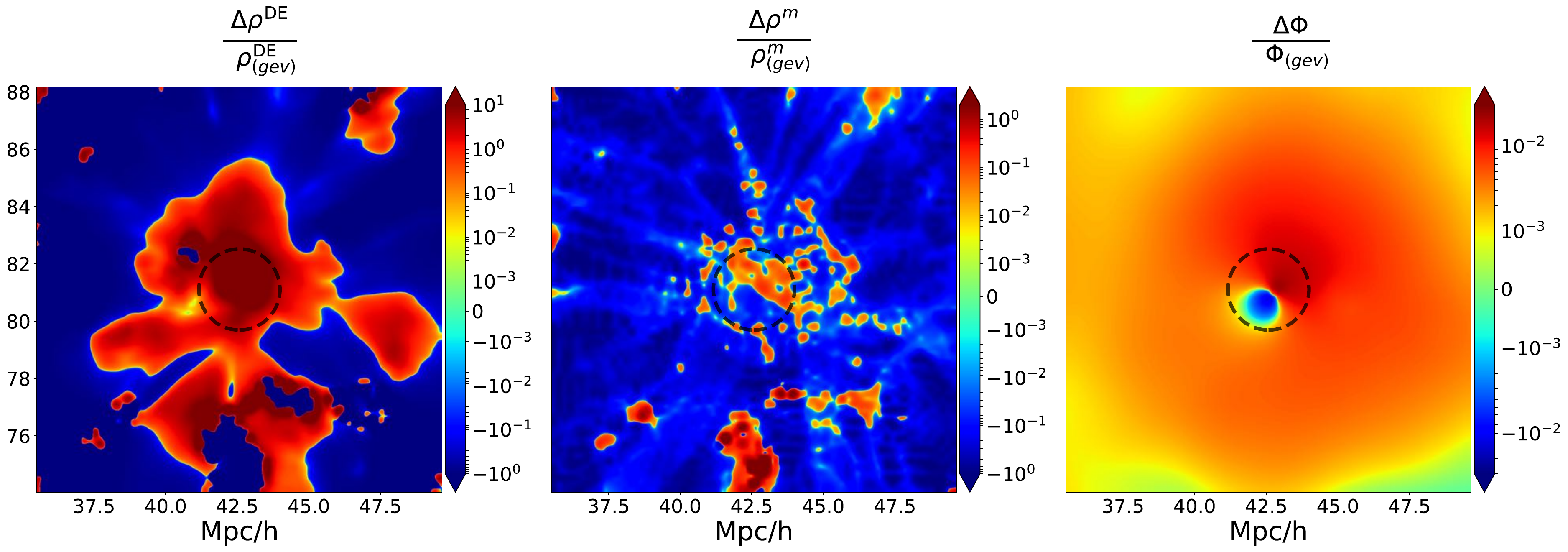} }
 \caption{The relative difference of dark energy density (left), matter density (middle) and gravitational potential (right) between the results from \kev and \gev at $z=0$.}
 \label{halo_rel_diff}
   \end{subfigure}
\caption{A comparison of dark energy clustering in simulations.}
 \end{figure}
 
For a more quantitative study we use higher-resolution simulations with 1024$^\mathrm{3}$ grid points and 100 Mpc$/h$ box size which corresponds to 0.097 Mpc$/h$ {spatial} resolution and {a mass resolution} of $8 \times 10^{8} \, \textup{M}_{\odot}/h$, for $c_s^2 =10^{-7}$ and $w=-0.9$. We pick the most massive halo in the simulation and analyse the particles and the $k$-essence scalar field inside five virial radii of the halo. Fig.~\ref{halos_po} shows, respectively from left to right, dark energy density in \gev and \kev and matter density in \kev at $z=0$. In each snapshot the position of the halo with {5 virial radii} is shown as a shaded region. Figure \ref{halo_close_look} provides a closer look at the halo where the dashed circle is the virial radius of the halo.
In \kev, dense dark energy structures are formed around the centre of the massive dark matter halo. In  Fig.~\ref{halo_rel_diff} the relative difference of dark energy density, matter density and potential between \kev and \gev is shown respectively on the left, middle and right, at $z=0$ and in the same region as Fig.~\ref{halo_close_look}. Due to the dark energy clustering that is absent in the linear realisation, we find a large change in the dark energy density distribution. Moreover, in contrast to matter power spectrum there are relatively large changes visible also in the matter density due to the dark energy non-linearity. The dipole visible in the distribution of the gravitational potential comes probably from a small shift of the halo center due to the dark energy non-linearity.

\begin{figure} [hbt!]
 \centering
 \includegraphics[scale=0.55]{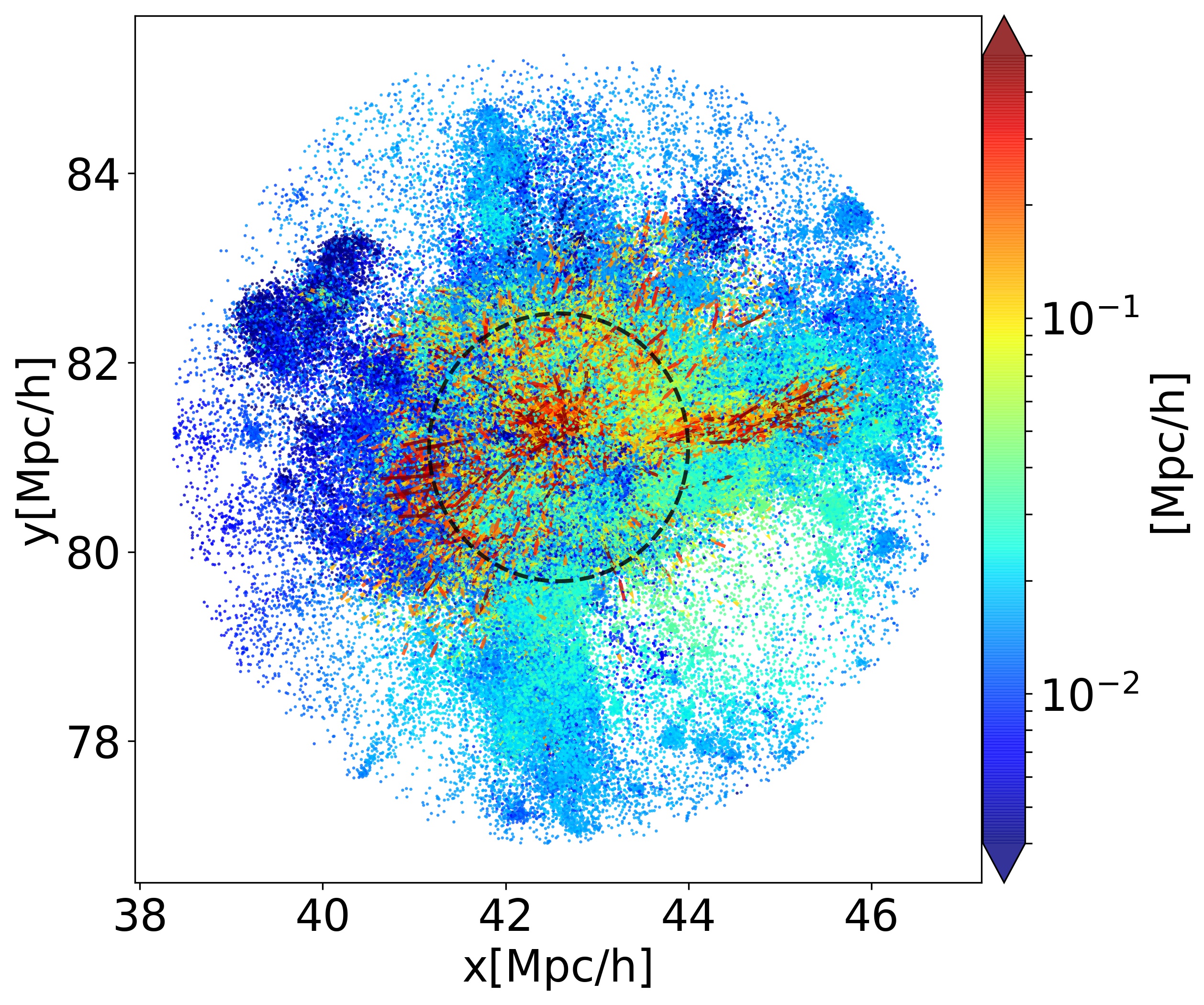} 
  \caption{Change in the particles positions due to  dark energy clustering. The arrows show the difference between the position of each particle in \kev and that of the same particle in \gev. The colors show the length of the arrows measured in $\rm Mpc/h$. Red arrows around the virial radius of the halo point toward the center.}
 \label{halo_pcls}
\end{figure}

In  Fig.~\ref{halo_pcls} the change in the position of particles in \kev with respect to the same particles in \gev inside three virial radii of the halo is shown.
Each arrow represents the displacement of the corresponding particle due to the dark energy non-linearities, i.e., $\Delta \vec r^{\,(i)} =\vec{r}^{\,(i)}_{\rm kevolution} - \vec {r}^{\, (i)}_{\rm gevolution}$, 
where  $\vec{r}^{\,(i)}_{\rm kevolution}$ is the position of particle $i$ in \kev and $\vec{r}^{\, (i)}_{\rm gevolution}$ is the position of the same particle in \gev. The colors show the length of the arrow measured in Mpc$/h$. Most changes in particles positions due to the dark energy non-linearities are seen to be around the center of the halo (for a study of the effect of dark energy clustering on the turn-around radius near galaxy clusters see \cite{Hansen:2019juz}).


\section{Conclusions}
 We develop \kev, an $N$-body code to compute cosmological observables including the effect of dark energy clustering. The code is based on \gev while 
dark energy is modelled using the EFT of DE. For simplicity we focus on $k$-essence but we pave the way to more general cases. We develop the equations to describe the gravitational and dark energy sector in 
the weak-field expansion but fully non-linearly. As a first initial step, however, we implement in the code only the linear parts of the evolution equation and stress-energy tensor of dark energy.

We compare the power spectra computed with \kev with those computed with codes that treat the evolution of the dark energy linearly, in particular with \gev 1.2 (where the dark energy stress-energy tensor is computed using \class) and with \class. 
We find relatively small differences between the matter power spectra computed with $k$-evolution and \gev. However, the clustering of dark energy uniquely captured by $k$-evolution affects non-negligibly the power spectra of other quantities, such as the gravitational potential and its time evolution. This is especially the case for low speeds of sound such as the one considered here, i.e.~$c_s^2 = 10^{-7}$.

Moreover, we compare \kev with simulations that take into account the dark energy component by changing the background evolution and back-scaling the initial conditions. 
We show that this back-scaling approach cannot compute, with sufficient accuracy, simultaneously the power spectrum of matter and of the gravitational potential. We also analyse snapshots from \kev. We  find that in dense regions the matter density, the $k$-essence density and the positions of particles are affected by dark energy clustering. 

This paper is the first step of a more general program of developing simulations including the effect of dark energy and modified gravity.

\label{sec:Conclusion}
\setcounter{equation}{0}

\section*{Acknowledgements}
 FH thanks Jean-Pierre Eckmann for helpful comments about manuscript and useful discussions. FH also would like to thank Benjamin L'Hu{i}llier, Arman Shafieloo, Pan Shi, Peter Wittwer, Mona Jalilvand, Hans Winther, Ruth Durrer, Lucas Lombriser, Miguel Zumalac{\'a}rregui, Eiichiro Komatsu, David Daverio, Joyce Byun, Yan-Chuan Cai and Matteo Cataneo for helping at different stages of the project and useful discussions.\\
This work was supported by a grant from the Swiss National Supercomputing Centre (CSCS) under project ID s710.
We acknowledge financial support from the Swiss National Science Foundation. {JA also acknowledges  funding by STFC Consolidated Grant ST/P000592/1.} Symbolic and numerical computations have been performed with MapleTM 16 and Wolfram Mathematica 9. {F.V.~acknowledges the Munich Institute for Astro- and Particle Physics (MIAPP) of the DFG cluster of excellence "Origin and Structure of the Universe”, where part of this work was done.}


 \appendix 
\section{Stress-energy  conservation and the $\pi$ equation of motion \label{sec:conseq}}

The stress-energy tensor for a perfect fluid reads,
\be
T^{\mu}_{\nu} = (\rho + p )u^{\mu} u_{\nu} + p \delta^{\mu}_{\nu} \;,
\ee
where $u^{\mu} $ is the fluid four-velocity normalized to unity, $u^\mu u_\mu =- 1$. This can be decomposed as
\be
u^{\mu} = \frac{dx^{\mu}}{ds} = \frac{e^{-\Psi}}{a}(1, v^i) \;,
\ee
so that the fluid components read
\be
\begin{split}
\label{SEcompsperfect}
T_0^0 &=  -\rho - \delta \rho \;, \\
T^i_0 &=-  (\rho + p) v^i = - e^{2 (\Phi + \Psi)} T^0_i \;, \\
T_i^j &= (p +\delta p ) \delta_i ^j + \Sigma_i^j  \;,
\end{split}
\ee
where $\Sigma_i^j\doteq T_i^j - \delta_i^j T^k_k/3 $  denote the anisotropic stress, which is traceless $\Sigma_i^i =0$.

The stress-energy tensor of a $k$-essence dark energy has the same form as the one of a perfect fluid. 
In particular, its components in \eqref{SEcomponents} can be written as  those of \eqref{SEcompsperfect} with  
\be
\label{SEfullfluids}
\begin{split}
\delta \rho & =  - \frac{\rho+p}{c_s^2} \bigg[ 3c_s^2 \HH \pi -\zetaf - \frac{2c_s^2 - 1}{2} (\vec{\nabla} \pi)^2   \bigg ] \;, \\
\delta p &  = -  \, (\rho+p)  \bigg[ 3 w  \HH \pi -\zetaf +  \frac16 (\vec{\nabla} \pi)^2   \bigg]  \;, \\
v^i & = - e^{2 (\Phi + \Psi)} \bigg[1- \frac{1}{c_s^2}   \, \Big( 3 c_s^2 (1+w) \mathcal{H} \pi -\zetaf + c_s^2 \Psi \Big) + \frac{c_s^2 - 1}{2 c_s^2}   (\vec{\nabla} \pi)^2  \bigg ] \partial_i \pi \;, \\
\Sigma_{ij} & = (\rho+p) \left[ \partial_i \pi \partial_j \pi - \frac13 (\partial_k \pi)^2 \delta_{ij} \right] \;.
\end{split}
\ee

For completeness, we use the covariant conservation of the stress-energy tensor of $k$-essence to derive the continuity and Euler equation of the dark energy fluid. Then we show that the continuity equation is equivalent to the equations of motion for $\pi$, Eqs.~\eqref{zeta_eq1} and \eqref{zeta_eq2}.
For convenience we define the following notation,
\be
\label{definitions}
w \doteq \frac{p}{\rho} \;, \qquad \delta \doteq \frac{\delta \rho}{\rho} \;, \qquad
\theta \doteq e^{-2(\Phi + \Psi)} \partial_i v^i \;,  \qquad \sigma \doteq \frac{\partial^{-2} \delta^{ik} \partial_k \partial_j \Sigma_{i}^j }{\rho+p} \;,
\ee
where $w$ is the equation of state parameter, $\delta$ is the density contrast and $\theta$ is the velocity divergence.
Note that for $k$-essence $v^i$ is irrotational, so that $\theta$ is enough to describe the full vector $v^i$.

If matter and dark energy are minimally coupled, as in the case of $k$-essence, the stress-energy tensor satifies the covariant conservation equation
\be
\nabla_\mu T^\mu_\nu =0 \;.
\ee
The continuity equation follows from taking this equation with $\nu=0$, which gives
\be
 \delta' = -(1+w) \big(\partial_i v^i- 3 \Phi' \big) -3 \mathcal{H}  \bigg(\frac{\delta p} {\delta \rho} -w\bigg) \, \delta + 3  \Phi'  \bigg( 1+ \frac{\delta p} {\delta \rho}  \bigg) \, \delta+ \frac{1+w}{\rho} v^i \partial_i  \big(3\Phi - \Psi \big) \;.
 \ee
With the above notation and keeping terms up to order ${\cal O}(\epsilon)$ in the weak-field expansion, one finds
\be
\label{conteq}
 \delta' = -(1+w) \big( \theta- 3 \Phi' \big) -3 \mathcal{H}  \bigg(\frac{\delta p} {\delta \rho} -w\bigg) \, \delta + 3  \Phi'  \bigg( 1+ \frac{\delta p} {\delta \rho}  \bigg) \, \delta+ \frac{1+w}{\rho} v^i \partial_i  \big(3\Phi - \Psi \big) \;.
 \ee

The Euler equation follows from $\nu=i$, which gives
\begin{align}
&  \rho  \, e^{-2 (\Phi + \Psi)} (v^{i})' (1+w) + e^{-2 (\Phi+\Psi)} (1+w) (3w-1) \mathcal{H} \rho \, v^{i} + \Sigma_i^j  \, \partial_j ( \Psi -3 \Phi)
\nonumber  \\ &
+ \partial_i \delta p  + \delta p  \, \partial_i \Psi 
- e^{-2 (\Phi + \Psi)} \rho v^i (1+w)  \,( 5 \Phi' +   \Psi')
+ \rho \, \partial_i \Psi (1+w) + \partial_i \Psi \,  \rho \, \delta + \partial_j \Sigma_i^j =0 \;.
\end{align}
Dividing this equation by $(1+w) \rho$, taking its divergence of  and replacing $\partial_i v^i$ using the definition of $\theta$ in Eq.~\eqref{definitions}, one finds  
\be
\begin{split}
&  \theta'+   (3w-1) \mathcal{H}  \, \theta +  \nabla^2 (\Psi + \sigma)+ \frac{\nabla^2 \delta P}{\rho(1+w)}  - ( 5 \Phi' +   \Psi') \theta +  \frac{\nabla^2 \Psi }{1+w}\bigg(1+  \frac{ \delta P } {\delta \rho}  \bigg) \, \delta
\\  &
- \frac{ \partial_ i  \Sigma_i^j}{\rho (1+w)} \partial_j ( {3  \Phi }  -   \Psi  )=0 \;.
\end{split}
\ee

One can verify that the continuity equation for the $k$-essence fluid is equivalent to field equations for $\pi$. 
We do it explicitly in the {limit of small speed of sound}, where 
\be
\label{smallcslimit}
\begin{split}
\delta & =  \frac{1+w}{2 c_s^2} \Big[2 \zetaf -   (\vec{\nabla} \pi)^2  \Big ] \;, \\
v^i &= - \frac{e^{2 (\Phi + \Psi)} } {2 c_s^2} \Big[2 \zetaf -   (\vec{\nabla} \pi)^2  \Big ] \partial _i \pi \;, \\
\delta p &=  - {\rho} \, (1+w)  \bigg[ 3 w  \HH \pi -\zetaf +  \frac16 (\vec{\nabla} \pi)^2   \bigg]  \;.
\end{split}
\ee
From the expression for $v^i$ and upon use of Eq.~\eqref{zeta_eq1},  the velocity divergence reads
\be
\theta= \frac{ 1}{2 c_s^2} \Big [ -2 \partial_i \zetaf \partial_i \pi - 2 \zetaf \nabla^2 \pi+ \partial_i \left( \partial_i \pi (\vec \nabla \pi)^2 \right) \Big] \;.
\ee
Putting the above expressions in the continuity equation \eqref{conteq} and multiplying by $c_s^2$, gives,
\be
 \zetaf' - 3 w  \HH \zetaf   -  \vec  \nabla \left( 2   \zetaf  +    \Psi \right) \cdot \vec  \nabla \pi - \zetaf \nabla^2 \pi     +  \frac{\HH (2+ 3 w)}{2}  (\vec \nabla \pi)^2       +\frac{1 }{2} \partial_i \left( \partial_i \pi (\vec \nabla \pi)^2 \right) =0 \;,
\ee
which is Eq.~\eqref{zeta_eq2} in the small $c_s^2 $ limit.

\section{Numerical implementation\label{sec:numimplement}}
We use the Newton-Stormer-Verlet-leapfrog method \cite{EHairer} to solve the two first order partial differential equations for the linear $k$-essence scalar field on the lattice,
 \be
  \zetaf^{\mf n+\frac{1}{2}}_\lat=  \zetaf^{\mf n-{\frac{1}{2}} } _\lat+  \zeta'^{ \; \mf n}_\lat \; \Delta \tau
  \ee
where the superscript ${\mf n}$ and subscript $\lat$ shows respectively the time step and the position on the lattice, i.e $ \zetaf^{\mf n+\frac{1}{2}}_\lat $ is the field $\zeta$ at discrete time step $(\mf n+\frac{1}{2})$ and point $(\lat)$  on the lattice.  To find $\zeta'^{ \; \mf n}_\lat$ {we discretize Eq.\eqref{zeta_eq2} as} 
\begin{align}
\begin{split}
 \zeta'^{ \;\mf n}_\lat  =&   3w \HH^{\mf n} \zetaf^{\mf n}_\lat  -  3 c_s^2 \HH^{\mf n} \left( \HH^{\mf n} \pi^{\mf n}_\lat  -  \Psi^{\mf n}_\lat - \frac{\HH'^{\mf n}}{\HH^{\mf n}} \pi^{\mf n}_\lat  - \frac{\Phi'^{\; \mf n}_\lat}{\HH^{\mf n}}  \right)  \nonumber \\  & +c_s^2  \frac{\Phi_{\mathrm{i}-1, \mathrm{j}, \mathrm{k}}^{\mathrm{n}}+\Phi_{\mathrm{i}+\mathrm{j}, \mathrm{k}}^{\mathrm{n}}+\Phi_{\mathrm{i}, \mathrm{j}-1, \mathrm{k}}^{\mathrm{n}}+\Phi_{\mathrm{i}, \mathrm{j}+1, \mathrm{k}}^{\mathrm{n}}+\Phi_{\mathrm{i}, \mathrm{j}, \mathrm{k}-1}^{\mathrm{n}}+\Phi_{\mathrm{i}, \mathrm{j}, \mathrm{k}+1}^{\mathrm{n}}-6 \Phi_{\mathrm{i}, \mathrm{j}, \mathrm{k}}^{\mathrm{n}}}{\mathrm{\Delta x}^{2}}
\end{split}
\end{align}
To update the scalar field {fluctuation} $\pi^{\sn}_\lat$ we use,
\be
      \pi^{\sn +1}_\lat=\pi^{\sn }_\lat + \pi'^{\; \sn+\frac{1}{2}}_\lat \Delta \tau \label{B2}
\ee
while $ \pi'^{\; \sn+\frac{1}{2}}_\lat$ is obtained by,
     \be
  \pi'^{\; \sn+\frac{1}{2}}_\lat=  \zetaf^{\sn+\frac{1}{2}}_\lat  - \mathcal{H}^{\sn+\frac{1}{2} }_\lat    \pi^{\sn+\frac{1}{2}}_\lat  +\Psi^{\sn+\frac{1}{2}} _\lat \label{B3}
      \ee
      It is important to note that in our scheme we have split the background from perturbations, as a result we have access to the $\mathcal{H}^{\sn +\frac12}$ independently of the value of the fields.
Moreover we compute $ \pi^{\sn+\frac{1}{2}}_\lat$ as following,
  \be
   \pi^{\sn+\frac{1}{2}} _\lat= \frac{ \pi^{\sn+1}_\lat +  \pi^{\sn}_\lat }{2 } \label{B4}
  \ee
 Putting \eqref{B4} and \eqref{B3} into the \eqref{B2}  results in,
 \be
 \pi^{\sn+1}_\lat=  \frac{1}{1+ \mathcal{H}^{\sn+\frac{1}{2}}  \Delta \tau/2}\Bigg[ \pi^{\sn}_\lat + \Delta \tau \Big [  \zetaf^{\sn+\frac{1}{2}} _\lat  -\mathcal{H}^{\sn+\frac{1}{2}}_\lat   \frac{  \pi^{\sn}_\lat }{2 } +\Psi^{\sn+\frac{1}{2}}  \Big ] \Bigg]
 \ee
where,
   \begin{align}
      \Psi^{\sn+\frac{1}{2}} _\lat= \Psi^{\sn}_\lat + \Psi'^{ \; \sn}_\lat \frac{\Delta \tau}{2} 
   \end{align} 
 Depending on the speed of sound of $k$-essence field $c_s^2$, we choose the appropriate Courant factor. Usually the $k$-essence Courant factor is different from the dark matter Courant factor and shows how many times the $k$-essence field is updated for one dark matter update. The reason is that for large speed of sound we need to decrease the time step of $k$-essence field updates to resolve the perturbations well.

\section{Initial conditions and gauge transformations}
\label{sec:IC}
In this appendix we discuss  the initial conditions for the scalar fluctuations $\pi$ and $\zeta$ in \kev, provided by the Boltzmann codes at high redshifts, where the linear theory is a good approximation.

In the linear Boltzmann code  \class,  dark energy is implemented as a fluid in both Newtonian and Synchronous gauge. In  \hiclass, $k$-essence is implemented in the field language in Synchronous gauge only. To extract $\pi$ and $\zeta$ from these codes, we need to gauge transform the perturbations to the Poisson gauge used in this article.

Let us first connect the field quantities to the fluid ones, in Newtonian gauge. To do that, we use the density contrast and velocity divergence of dark energy, respectively $\delta$ and $\theta$, at linear order given in Eq.~\eqref{SEfullfluids}. Using these expression we  find
\begin{align}
\pi_\mf{Newt}(k,z) & =\frac{\theta_\mf{Newt} (k,z)}{k^2}\;, \\ 
\pi'_{\mf {Newt}}(k,z) &= \frac{c_s^2}{1+w} \delta_{\mf{Newt}} (k,z) + c_s^2 \mathcal{H} \frac{\theta_{\mf{Newt} }(k,z)}{k^2} +\Psi_{\mf{Newt} }(k,z) \;,
\label{theta_pi}
\end{align}
where the subscript ``$\mf{Newt}$'' denotes conformal Newtonian gauge.

Thus,  using Eq.~\eqref{pi_eq} we then obtain, for $\zeta$,
\be
\zetaf_{\mf{Newt} } =  \frac{c_s^2}{1+w} \delta_{\mf{Newt} }(k,z) + \mathcal{H} \frac{\theta_{\mf{Newt} }(k,z)}{k^2} (1+ c_s^2) \;.
\ee
Following the discussion in Sec.~3 of \cite{Ma:1995ey} on the gauge transformations and employing the same notation, we consider the following coordinate transformation:
\begin{align}
 {\hat{x}}^0 &= x^ 0 + \alpha ,  \nonumber \\ 
{\hat{\vec{x}}} &={\vec{x}} +\vec{\nabla} \beta(\tau,x) + \vec{\epsilon } (\tau,x), \; \; \; \; \; \vec{\nabla} . \vec{\epsilon}=0
\end{align}
where $\alpha$ and $\beta$ are respectively the temporal and spatial part of the infinitesimal coordinate transformation. The metric components transform as,
\be
{\hat{g}_{\mu \nu }}  (x)=  g_{\mu \nu } (x) - g_{\mu \beta } (x) \partial _{\nu} \epsilon^{\beta} -  g_{\alpha \nu } (x) \partial _{\mu} \epsilon^{\alpha}-\epsilon^{\alpha} \partial _{\alpha} g_{\mu \nu } (x) \;.
\ee
As a result the transformed metric perturbations after coordinate transformation read,
\begin{align}
&\hat {\Psi} (\tau,\vec{x}) = \Psi(\tau,\vec{x}) - \alpha' (\tau,\vec{x})- \mathcal{H} \alpha(\tau,\vec{x}) \nonumber \;, \\ &
\hat {\Phi} (\tau,\vec{x}) = \Phi(\tau,\vec{x}) +\frac{1}{3} \nabla^2 \beta(\tau,\vec{x})  + \mathcal{H} \alpha(\tau,\vec{x}) \;, 
\end{align}
where we have assumed that the coordinate transformation is of the same order as the metric perturbations.
We can use these transformations to write down the scalar metric perturbations in Newtonian gauge $(\Phi,\Psi)$ in terms of $(h,\eta)$ in Synchronous gauge. The result is that in Fourier space we can set
\be
\alpha (\tau,k)=({h'} + 6 {\eta'})/2 k^2, \qquad
\beta(\tau,k) =({h} + 6 {\eta})/2 k^2 \;,
\ee
where $(h,\eta)$ are the scalar modes of $h_{ij}$ and are defined as,
\be 
h_{i j}^{ \|}(\vec{x}, \tau)=\int d^{3} k e^{i \vec{k} \cdot \vec{x}}\left(\hat{k}_{i} \hat{k}_{j}-\frac{1}{3} \delta_{i j}\right)\{h(\vec{k}, \tau)+6 \eta(\vec{k}, \tau)\}, \qquad \vec{k}=k \hat{k} \, .
\ee

 To find the gauge transformation for the fluid quantities we use
\be
T^{\mu} _ {\nu}(\mf{Synch}) = \frac{\partial {\hat{x}} ^{\mu} }{\partial {x} ^{\sigma}} \frac{\partial {x} ^{\rho}}{\partial {\hat{x}} ^{\nu}} T^{\rho} _ {\sigma}(\mf{Newt}) 
\ee
where ${\hat{x}}^{\mu}$ and $x^{\mu}$ denote the Synchronous and Newtonian coordinates respectively.
It follows, to linear order, that
\begin{align}
& T_0^0 (\mf{Synch}) =T_0^0 (\mf{Newt}) \, ,\\
\nonumber &
  T_0^j (\mf{Synch}) =T_0^j (\mf{Newt}) + i k_j \alpha (\bar{\rho} + \bar{P}) \, , \\
\nonumber &
   T_i^j (\mf{Synch}) =T_i^j (\mf{Newt}) \, .
 \end{align} 
 
From the definitions of the density contrast $\delta = \delta \rho /\bar{\rho} = - \delta T_0^0 /\bar{\rho} $, $\theta$, $\delta P$ and $\sigma$ in Eq.~\eqref{SEfullfluids} we have,
\begin{align}
\delta (\mf{Synch}) &= \delta (\mf{Newt}) -\alpha \frac{\dot{\bar{\rho}}}{\rho} \, ,
 \\
\nonumber  
\theta (\mf{Synch}) &= \theta (\mf{Newt}) - \alpha k ^2 \, , \\
\nonumber 
\delta P (\mf{Synch}) &= \delta P (\mf{Newt}) - \alpha \dot{\bar{P}} \, , \\ 
\nonumber 
\sigma  (\mf{Synch}) &= \sigma  (\mf{Newt}) \, .
\end{align}

To obtain the gauge transformation for $\pi$ we use the fact that $\varphi  (x^{\mu} ) =x^0 + \pi (x^{\mu})$ is a scalar, i.e.,~$\hat{\varphi }({\hat{x}}^{\mu})=\varphi ({x}^{\mu})$. Thus, we find
\begin{align}
\Delta \varphi (x) & \doteq \hat{ \varphi}(x) - \varphi (x) 
= - \partial _ {\mu} \varphi \, \epsilon ^{\mu}
 = \epsilon ^0 = -\alpha  \, ,
\end{align}

which can be written as
\begin{align}
\pi_{\mf{Synch}} &=\pi_{\mf{Newt}} - \alpha  \,.
\end{align}

\begin{figure} [H]
\centering
 \includegraphics[scale=0.26]{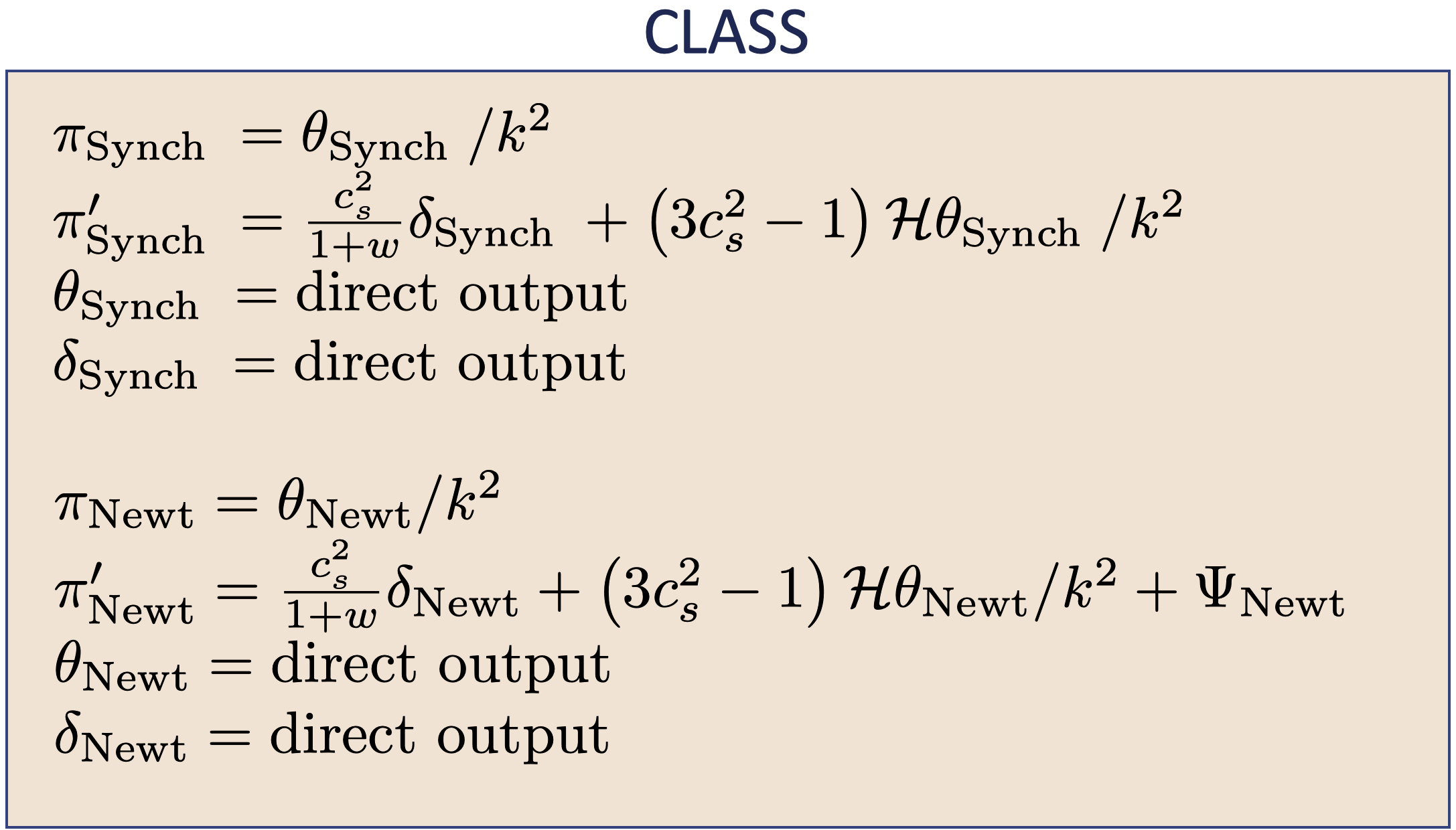} 
 \caption{{The transformations in \class to obtain $\pi$ and $\pi'$ in a certain gauge from the fluid properties. In the top part, it is assumed that the user runs \class in Synchronous gauge which $\theta_{\mf{Synch}}$ and $\delta_{\mf{Synch}}$ in Synchronous gauge are the direct output of the code. To obtain $\pi_{\text{Synch}}$ and $\pi'_{\text{Synch}}$ one needs to use the given transformations. In the bottom part, it is assumed that the user runs \class in Newtonian gauge. Follow the recipe one obtains $\pi_{\text{Newt}}$ and $\pi'_{\text{Newt}}$}}.
 \label{transformation_class}
 \end{figure}

\begin{figure} [H]
\centering
 \includegraphics[scale=0.26]{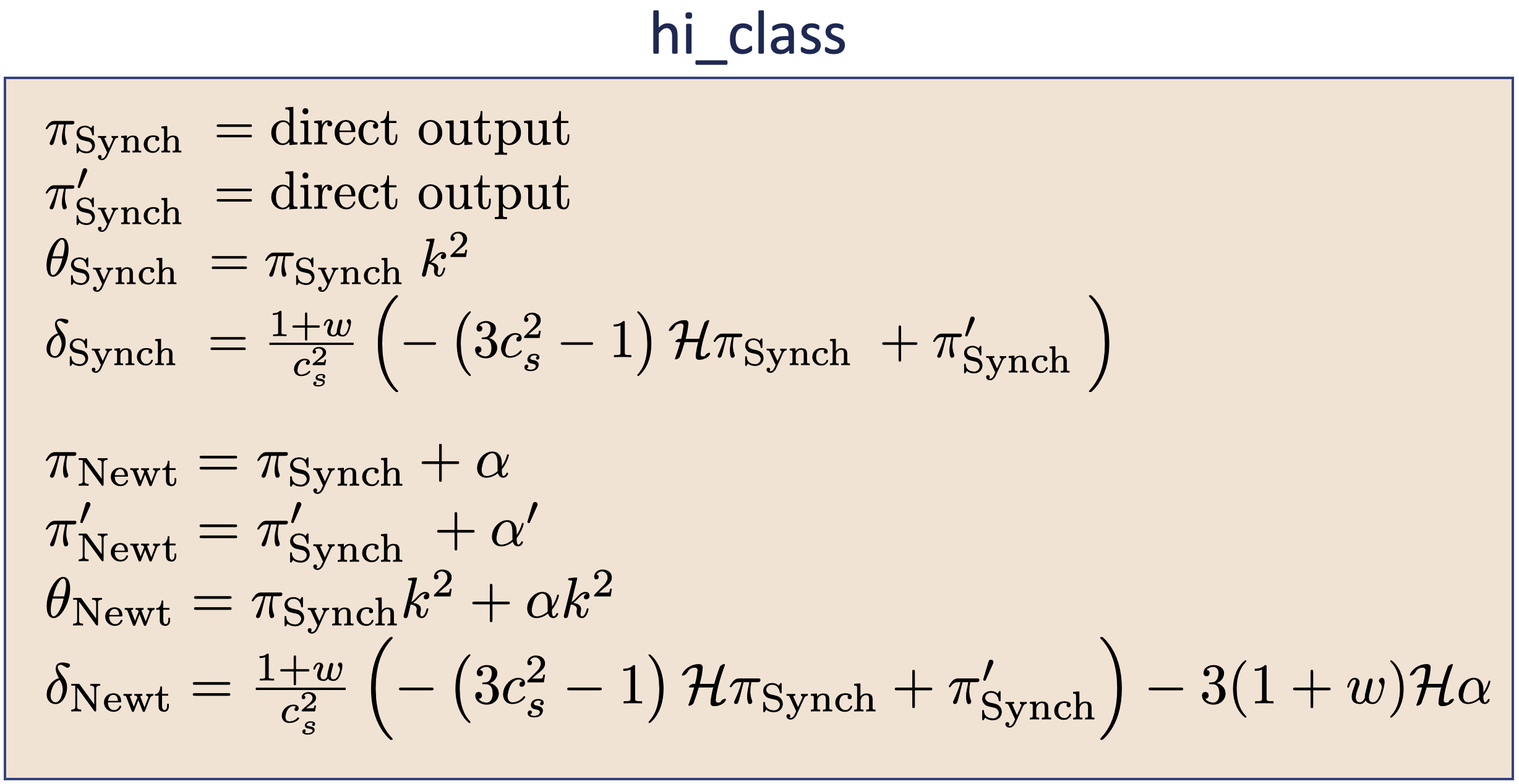} 
 \caption{{ The transformations in \hiclass to obtain a certain quantity in a specific gauge. In the top, the recipe in synchronous gauge is given. In \hiclass, $\pi_{\text{Synch}}$ and $\pi'_{\text{Synch}}$ are the direct output of the code. To obtain $\delta_{\text{Synch}}$ and $\theta_{\text{Synch}}$ one needs to follow the given transformations. In the bottom part, the recipe for obtaining quantities in Newtonian gauge in both languages are given.}}
 \label{transformation_hiclass}
 \end{figure}

For practical applications we provide a list of transformations in Fig.~\ref{transformation_class} and Fig.~\ref{transformation_hiclass} for the output of \class and \hiclass to obtain the result in a certain language with a certain gauge.
In \class, we assume that the user sets the correct gauge, for example to have the quantities in Synchronous gauge one uses \class with the gauge is set to ``Synchronous''. Then $\delta$ and $\theta$ in that gauge are the output of the code and one has to use the transformations in Fig.~\ref{transformation_class} to find $\pi$, $\pi'$ in the corresponding gauge. 
 
 In \hiclass the quantities are written in Synchronous gauge only, in the field language. In this case $\pi_{\mf{Synch}}$ and $\pi'_{\mf{Synch}}$ are the outputs of the code, while $\delta_{\mf{Synch}}$,  $\theta_{\mf{Synch}}$, $\pi_{\mf{Newt}}$ and $\pi'_{\mf{Newt}}$ are computed according to the formulas given in
Fig.~\ref{transformation_hiclass}.

\section{{Limit of small speed of sound $c_s^2$} }
\label{sec:smallcslimit}

In this appendix we are going to study the evolution of perturbations in the limit of small speed of sound, in order to show that this limit is well defined.  In particular, we are going to show that $\delta$ and $v^i$ remain finite in this limit, despite the appearance of a $c_s^2$ in the denominator of their expressions, see Eq.~\eqref{smallcslimit}.
To be able to solve this case analytically, we will assume matter dominance, i.e., $a \propto \tau^2$ (i.e.,~$\HH = 2/\tau$) and $\Psi =$ const.
Moreover, we expand $\zeta$ and $\pi$ in perturbations, 
\be
\zetaf = \zetaf^{(1)} + \zetaf^{(2)} + \ldots \;, \qquad \pi = \pi^{(1)} + \pi^{(2)} + \ldots  \;,
\ee
and we start by discussing linear perturbations.

At first order the evolution equations for $\pi$ read
\be
\label{linearEOM}
\begin{split}
&\zetaf - \pi ' - \HH \pi + \Psi =0 \;, \\
 &\zetaf' - 3 w \HH \zetaf+ 3 c_s^2 \left(\HH^2 \pi - \HH \Psi - \HH' \pi - \Phi' \right) -c_s^2 \nabla^2 \pi =0 \;,
\end{split}
\ee
where  we initially omit the upper index ${(1)}$ to avoid cluttering.
We can solve these equations for $\zetaf$ and $\pi$ perturbatively in $c_s^2$, i.e.,~using the expansions
\be
\zetaf = \zetaf_0 + \zetaf_1 c_s^2 + \ldots \;, \qquad \pi = \pi_0 + \pi_1 c_s^2+ \ldots \;.
\ee

At lowest order in $c_s^2$,  the second equation becomes
\be
 \zetaf_0' - 3 w \HH \zetaf_0=0 \;.
\ee
Assuming $w$ constant, its solution reads
$ \zetaf_0  = C \, a^{-3w}$,
where $C$ is an arbitrary constant.
However, in order to prevent the stress-energy tensor of dark energy from blowing up for $c_s^2 \to 0$ (see Eq.~\eqref{smallcslimit}), we  fix it to zero, $C=0$, so that $\zetaf$ starts at linear order in $c_s^2$, 
\be
\zetaf_0^{(1)} =0\;.
\ee
Notice that this coincides with assuming adiabatic initial conditions, i.e., $\Phi =  \dot \xi^0 = \pi' + \HH \pi $ \cite{Gleyzes:2014rba}. Plugging $\zetaf_0=0$ in the first equation of Eq.~\eqref{linearEOM} we can solve for $\pi$,
\be
\label{linearpi}
 \pi_0^{(1)}=  \frac{ \Psi }{3 } \tau \;.
\ee

Since $\zeta$  vanishes at leading order in $c_s^2$, let  us go to the next order. At first order in $c_s^2$, the second equation in Eq.~\eqref{linearEOM} reads
\be
 \zetaf'_1 - 3 w \HH \zetaf_1+ 3\left(\HH^2 \pi_0 - \HH \Psi - \HH' \pi_0 - \Phi' \right) - \nabla^2 \pi_0 =0 \;.
\ee
Using the  solution for $\pi$ and taking $\Phi'=0$,  we can solve for $\zetaf_1$,  which gives
\be
 \zetaf_1^{(1)}  = \frac{ \tau^2 } {6 (1-3w)}  \nabla^2 \Psi \;.
   \label{eqzeta}
\ee

Since at leading order in  $c_s^2$ we have $\delta =({1+w}) \zetaf /{c_s^2} $,
see Eq.~\eqref{smallcslimit},  and in matter domination the gravitational potential is given by the matter density contrast $\delta_{\rm m}$ by  the usual Poisson equation,
$\nabla^2 \Psi  = ({3}/{2}) \mathcal{H}^2 \delta_{\rm m}  $,
the above solution for $\zetaf^{(1)} = \zetaf_1^{(1)} c_s^2$ gives
\be
\delta^{(1)}  = \frac{1+w}{1-3w}  \,\delta_{\rm m}^{(1)} \;,
\ee
as expected \cite{Creminelli:2008wc}.

At second order in perturbations  we have
\be
 \zetaf_0^{(2)}{}' - 3 w \HH \zetaf_0^{(2)}-  \vec  \nabla   \Psi  \cdot \vec  \nabla \pi_0^{(1)}     +  \frac{\HH}{2} (2+ 3 w  ) \big(\vec \nabla \pi_0^{(1)} \big)^2   =0 \;,
\ee
where  in the second equation we have assumed $c_s^2 =0 $ and have taken only the leading order  in the expansions $\zetaf = \zetaf_0 + \zetaf_1 c_s^2 + \dots $ and $\pi = \pi_0 + \pi_1 c_s^2 + \ldots $.
Using the linear solution for $\pi$, this equation can be solved,
\be
 \zetaf_0^{(2)}   =  \frac{\tau^2 }{18 } ( \vec  \nabla    \Psi )^2 \;,
\ee
which is exactly what needed to cancel the right-hand side of the first two equations in Eq.~\eqref{smallcslimit}.
To solve for $\pi_0^{(2)}$ we use that  $\Psi^{(2)}=0$ in Eq.~\eqref{zeta_eq1}, which gives
\be
 \pi_0^{(2)} =    \frac{\tau^3 }{90 } ( \vec  \nabla    \Psi )^2   \;.
\ee
In this case we do not need to go one order higher in $c_s^2$ to find  $\zeta$, because the leading order does not vanish. 
Using Eq.~\eqref{smallcslimit}, the above solutions show that $\delta^{(2)}$ and $v_i^{(2)}$ vanish at this order in $c_s^2$. To find these quantities at leading order in $c_s^2$ we need to solve for $\zetaf_1^{(2)}$ and $\pi^{(2)}$ (which requires $\pi_1^{(2)}$) and replace these quantities in Eq.~\eqref{SEfullfluids}.  
Since the  solution obtained by this straightforward procedure is not very illuminating, we refrain from giving it here.

Going one order higher in perturbations, 
at third order the evolution equation of $\zetaf_0$ reads
\be
\begin{split}
& \zetaf_0^{(3)}{}' - 3 w \HH \zetaf_0^{(3)}-2 \vec  \nabla \zetaf_0^{(2)} \cdot \vec  \nabla \pi_0^{(1)} - \vec \nabla \Psi \cdot \vec \nabla \pi_0^{(2)} -\zetaf_0^{(2)}     \nabla^2 \pi_0^{(1)}    +  {\HH} (2+ 3 w  ) \vec \nabla \pi_0^{(2)} \cdot  \vec \nabla \pi_0^{(1)} 
\\ &+ \frac1 2 \nabla_i \left[ \nabla_i \pi^{(1)}_{0} \Big( \vec \nabla \pi^{(1)}_0 \cdot  \vec \nabla \pi^{(1)}_{0} \Big) \right]  =0 \;.
\end{split}
\ee
Using the first and second-order solutions for $\pi$ written above, this equation can be solved giving
\be
 \zetaf_{0}^{(3)}    = \frac{\tau^4}{270  } \vec \nabla \Psi \cdot \vec \nabla (\vec \nabla \Psi)^2  \;. 
\ee
Replacing this solution  in Eq.~\eqref{smallcslimit} with the lowest order solutions for $\pi$ shows again that  $\delta$ remains finite in the $c_s^2 \to 0$ limit. This procedure can be straightforwardly extended to higher orders.

The solution for $\pi_0^{(3)}$ can be found by solving $\zetaf_0^{(3)} - \pi_0^{(3) }{}'  - \HH \pi_0^{(3)} =0$, giving
\be
  \pi_0^{(3) }   = \frac{ \tau^5}{1890} \vec \nabla \Psi \cdot \vec \nabla (\vec \nabla \Psi)^2  \;.
  \ee
In summary, at leading order in $c_s^2$ and up to third order in perturbations we  have
\be
\begin{split}
 \zetaf &= \frac{ \tau^2 c_s^2 } {6 (1-3w)}  \nabla^2 \Psi  +     \frac{\tau^2 }{18 } ( \vec  \nabla    \Psi )^2 
 + \frac{\tau^4}{270} \vec \nabla \Psi \cdot \vec \nabla (\vec \nabla \Psi)^2   \;, \\
 \pi &=  \frac{\tau \Psi}{3}  + \frac{\tau^3 }{90 } ( \vec  \nabla    \Psi )^2  + \frac{\tau^5}{1890} \vec \nabla \Psi \cdot \vec \nabla (\vec \nabla \Psi)^2 \;.
 \end{split}
\ee

\bibliographystyle{JHEP}
\bibliography{Biblio_kevolution}

\end{document}